\documentclass[graybox]{svmult}

\usepackage{mathptmx}       
\usepackage{helvet}         
\usepackage{courier}        
\usepackage{type1cm}        
\usepackage{makeidx}         
\usepackage{graphicx}        
\usepackage{multicol}        
\usepackage[bottom]{footmisc}

\def\gapprox{$_>\atop{^\sim}$} \def\lapprox{$_<\atop{^\sim}$}
\def\nhi{\noindent \hangindent=10pt} 
\def\etal{{et~al.\ }}

\makeindex             

\begin{document}

\title*{Elliptical Galaxies and Bulges of Disk Galaxies: Summary of Progress and Outstanding Issues}

\titlerunning{Formation of Elliptical Galaxies and Bulges: Progress and Outstanding Issues} 

\author{\bf John Kormendy}

\institute{John Kormendy \at Department of Astronomy, University of Texas at Austin, 
           2515 Speedway, Mail Stop C1400, Austin, Texas 78712-1205, USA \quad \email{kormendy@astro.as.utexas.edu}}

\maketitle


\includegraphics{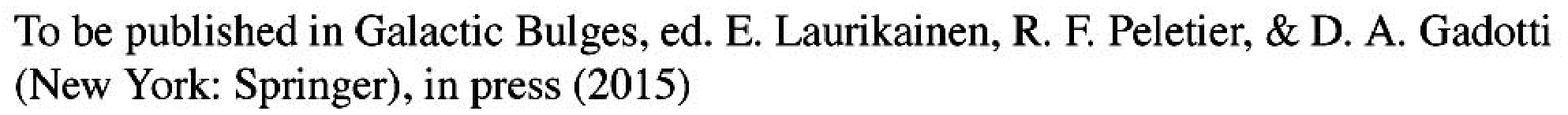}

\abstract*{Bulge components of disk galaxies are the high-density centers interior to their outer disks.  Once thought 
to be equivalent to elliptical galaxies, their observed properties and formation histories turn out to be richer 
and more varied than those of ellipticals.  This book reviews progress in many areas of bulge studies.
Two advances deserve emphasis:~(1) Observations divide bulges into ``classical bulges'' that 
look indistinguishable from ellipticals and ``pseudobulges'' that are diskier and (except in S0s) more actively 
star-forming than are ellipticals.  Classical bulges and ellipticals are thought to form by major galaxy mergers. 
Disky pseudobulges are a product of the slow (``secular'') evolution of galaxy disks.
Nonaxisymmetries such as bars and oval distortions transport some disk gas toward the center,
where it starbursts and builds a dense central component that is diskier in structure than are classical
bulges.  Secular evolution explains many regular structures (e.{\ts}g., rings) seen in galaxy disks.  It is a
new area of galaxy evolution work that complements hierarchical clustering. (2) Studies of 
high-redshift galaxies reveal that their disks are so gas-rich that they are violently unstable to the formation 
of mass clumps that sink to the center and merge.  \hbox{This is an alternative channel for the formation of classical bulges.}
\phantom{000} 
\phantom{000} 
This chapter summarizes big-picture successes and unsolved problems in the formation of bulges
and ellipticals and their coevolution (or not) with supermassive black holes.  I present an observer's perspective
on simulations of cold dark matter galaxy formation including baryonic physics.  Our picture of the quenching 
of star formation is becoming general and secure at redshifts $z < 1$.  I conclude with a list
of major uncertainties and problems.  The biggest challenge is to produce realistic bulges\ts$+${\ts}ellipticals
and realistic disks that overlap over a factor of $>$\ts1000 in mass but that differ from each other as we
observe over that whole range.  A related difficulty is how hierarchical clustering makes so many giant, bulgeless galaxies
in field but not cluster environments.  I present arguments that we rely too much on star-formation feedback and AGN feedback 
to solve these challenges.
\pretolerance=10000\tolerance=10000
}

\abstract{Bulge components of disk galaxies are the high-density centers interior to their outer disks.  Once thought 
to be equivalent to elliptical galaxies, their observed properties and formation histories turn out to be richer 
and more varied than those of ellipticals.  This book reviews progress in many areas of bulge studies.
Two advances deserve emphasis:~(1) Observations divide bulges into ``classical bulges'' that 
look indistinguishable from ellipticals and ``pseudobulges'' that are diskier and (except in S0s) more actively 
star-forming than are ellipticals.  Classical bulges and ellipticals are thought to form by major galaxy mergers. 
Disky pseudobulges are a product of the slow (``secular'') evolution of galaxy disks.
Nonaxisymmetries such as bars and oval distortions transport some disk gas toward the center,
where it starbursts and builds a dense central component that is diskier in structure than are classical
bulges.  Secular evolution explains many regular structures (e.{\ts}g., rings) seen in galaxy disks.  It is a
new area of galaxy evolution work that complements hierarchical clustering. (2) Studies of 
high-redshift galaxies reveal that their disks are so gas-rich that they are violently unstable to the formation 
of mass clumps that sink to the center and merge.  \hbox{This is an alternative channel for the formation of classical bulges.}
\phantom{000} 
\phantom{000} 
This chapter summarizes big-picture successes and unsolved problems in the formation of bulges
and ellipticals and their coevolution (or not) with supermassive black holes.  I present an observer's perspective
on simulations of cold dark matter galaxy formation including baryonic physics.  Our picture of the quenching 
of star formation is becoming general and secure at redshifts $z < 1$.  I conclude with a list
of major uncertainties and problems.  The biggest challenge is to produce realistic bulges\ts$+${\ts}ellipticals
and realistic disks that overlap over a factor of $>$\ts1000 in mass but that differ from each other as we
observe over that whole range.  A related difficulty is how hierarchical clustering makes so many giant, bulgeless galaxies
in field but not cluster environments.  I present arguments that we rely too much on star-formation feedback and AGN feedback 
to solve these challenges.
\pretolerance=10000\tolerance=10000
}

\section{Introduction}
\label{sec:1}
This final chapter summarizes areas of major progress in understanding galaxy bulges
and tries to distill the important unresolved issues that need further work. 

      I do not revisit the subjects covered by all chapters --
Madore (2015: historical review),
M\'endez-Abreu (2015: intrinsic shapes), 
Falc\'on-Barroso (2015: kinematic observations),
S\'anchez-Bl\'azquez (2015: stellar populations),
Laurikainen \& Salo (2015: observations of boxy bulges),
Athanassoula 2015: modeling of boxy bulges),
Gonzalez \& Gadotti (2015: observations of the Milky Way boxy bulge),
Shen \& Li (2015: modeling of the Milky Way boxy bulge),
Cole \& Debattista (2015: nuclear star clusters), and
Combes (2015: bulge formation within MOND).
I comment briefly on Zaritsky's (2015) chapter on scaling relations.

I concentrate in this summary chapter on three main areas of progress and on two main
areas where there are unresolved difficulties:

Two additions to our picture of bulge formation are (1) formation by massive clump instabilities 
in high-$z$ disks; Bournaud (2015) develops this story, but it deserves emphasis here, too, and 
(2) our picture of secular evolution of galaxy disks that produces two distinct kinds of dense central 
components in galaxies, disky pseudobulges (reviewed here by Fisher \& Drory 2015) and boxy pseudobulges
(discussed in four chapters listed above).  Both deserve emphasis here, too.

The main areas with unresolved issues come in two varieties: 

Probably the most important chapter in this book is Brooks\ts\&{\ts}Christensen (2015)
on the modeling of galaxy{\ts}--{\ts}and thus also bulge{\ts}--{\ts}formation.~These~models 
that add baryonic physics to giant $n$-body simulations of the hierarchical clustering of
cold dark matter (CDM) in a $\Lambda$CDM universe  define
the state of the art in the most general version of galaxy formation theory.  Much has
been accomplished, and progress is rapid.  Brooks \& Christensen (2015) is an excellent
review of the state of the art as seen by its practitioners. In this chapter, I 
would like to add the viewpoint of an observer of galaxy archaeology.  I suggest
a slightly different emphasis on the successes and 
shortcomings of present models.  My main purpose is to promote a dialog between theorists 
and observers that may help to refine the observational constraints that are most telling and 
the modeling exercises that may be most profitable.  Baryonic galaxy formation is an 
extraordinarily rich and difficult problem.  Many groups struggle honorably 
and carefully with different aspects of it.  In this subject, besides a strong push on 
remaining limitations such as resolution, the main need seems to me to be a broader
use of observational constraints and a consequent refinement of the physics that may succeed 
in explaining them.

A second issue involves Graham's (2015) chapter on supermassive black holes.
It is inconsistent with all other work that I am aware of on this subject, including
McConnell \& Ma (2013) and Kormendy \& Ho (2013).  Section 6 summarizes this subject
using results from Kormendy \& Ho (2013, hereafter KH13).  

Section 7 reviews the quenching of star formation in galaxies.  Many different lines of research
are converging on a consistent picture of how quenching happens.

Finally, I conclude with a personal view of the most important, big-picture issues that are 
still unsolved by our developing picture of galaxy evolution.

\section{Secular Evolution and the Formation of Pseudobulges}
\label{sec:2}

      Progress on bulge formation is dominated by two conceptual advances.  This section revisits
secular evolution in disk galaxies.  This is a major addition~that~complements our picture of galaxy 
evolution by hierarchical clustering.  I begin here because {\it all further discussion depends on the
resulting realization that the dense central components in galaxies come in two varieties with different 
formation processes, classical and pseudo bulges.}  Section 3 discusses the second conceptual advance,
the discovery of a new channel for the formation of classical bulges.  This is the formation at high $z$ 
of unstable clumps in gas-rich disks; they sink to the center along with lots of disk gas and starburst 
and relax violently.  In this way, bulge formation proceeds largely as it does during major mergers. 
This leads to a discussion of the merger formation of both bulges and ellipticals in Section\ts4.

      Our pictures of the merger formation of classical bulges and ellipticals~and~the secular growth of
pseudobulges out of disks both got their start in the late 1970s.  The importance of major mergers 
(Toomre \& Toomre 1972; Toomre 1977) in a hierarchically clustering universe (White \& Rees 1978) got a 
major boost from the realization that CDM halos make galaxy collision cross sections much bigger than they look.
This subject ``took off'' and rapidly came to control our formation paradigm.  Secular evolution is a 
more difficult subject -- slow processes are hard to study -- and it did not get a similar boost from the
CDM revolution.  However, the earliest papers on the subject come from the same time period: e.{\ts}g.,
Kormendy (1979a) emphasized the importance of slow interactions between nonaxisymmetric galaxy components;
Kormendy (1979b) first pointed out the existence of surprisingly disky bulges;
Combes \& Sanders (1981) showed that boxy pseudobulges are edge-on bars.
Kormendy (1981, 1982) reviewed and extended the results on disky bulges.
This subject did not penetrate the galaxy formation folklore; rather, it remained a series of active
but unconnected ``cottage industries'' for the next two decades.  Nevertheless, by the 1990s, the
concept{\ts}--{\ts}if not yet the name\ts-- of disky pseudobulges was well established (see Kormendy 1993 for a review),
and the idea that boxy bulges are edge-on bars was well accepted (see Athanassoula 2005 for a 
more recent and thorough discussion).  I hope it is fair to say that the comprehensive review by Kormendy 
\& Kennicutt (2004) has helped to convert this subject into a recognized paradigm -- it certainly is so in this 
book -- although it is still not as widely understood or taken into account as is hierarchical clustering.

      Kormendy \& Kennicutt (2004) remains up-to-date and comprehensive on the basic results and on observations
of prototypical pseudobulges.  However, new reviews extend and complement it.  Kormendy \& Fisher (2005, 2008)
and Kormendy (2008, 2012) provide the most important physical argument that was missing in Kormendy \& Kennicutt (2004):
Essentially all self-gravitating systems evolve toward more negative total energies (more strongly bound configurations)
by processes that transport kinetic energy or angular momentum outward.  In this sense, the secular
growth of pseudobulges in galaxy disks is analogous to the growth of stars in protostellar disks, the growth of
black holes in black hole accretion disks, the sinking of Jupiters via the production of colder
Neptunes in protoplanetary disks, core collapse in globular clusters, and the evolution of stars into red
(super)giants with central proto white dwarfs, \hbox{neutron stars, or stellar-mass} black holes.  All of these evolution 
processes are related.  So secular disk evolution and the growth of pseudobulges is very
fundamental, provided that some process redistributes angular momentum~in~the~disk.
My Canary Islands Winter School lectures (Kormendy 2012) are an \hbox{up-to-date} observational review that 
includes environmental secular evolution.  Sellwood (2014) provides an excellent theoretical review.

      Boxy pseudobulges are discussed in four chapters of this book; I concentrate on disky pseudobulges.
Fisher \& Drory (2015) review the distinction between classical and pseudo bulges from a
purely phenomenological point of view.~That~is,~they intercompare observational diagnostics to distinguish
between the two bulge types with no reference to physical interpretation.  This is useful, because
it gives relatively unbiased failure probabilities for each diagnostic.  They are not wholly 
independent, of course, because they are intercompared.  But they are independent enough in execution so that we get a
sufficient estimate of the failure probability when they are combined by multiplying the individual failure probabilities.  

      Kormendy \& Kennicutt (2004), Kormendy (2012), and KH13 strongly advocate the use of as
many bulge classification criteria as possible.  The reason is that any one criterion has a non-zero
probability of failure.  Confusion in the literature (e.{\ts}g., Graham 2011) results from the fact that
some authors use a single classification criterion (e.{\ts}g., S\'ersic index) and so get results that conflict
with those derived using multiple criteria.  But we have long known that most classical bulges have $n \geq 2$,
that most pseudobulges have $n < 2$, and that there are exceptions to both criteria.  No-one should be
surprised that S\'ersic index sometimes fails to correctly classify a bulge.  This is the point that
Fisher \& Drory (2015) make quantitative.

      Fisher \& Drory (2015) show that the failure probability of each classification criterion
that they test is typically 10\ts--\ts20\ts\%.  A few criteria are completely robust (if $B/T$ \gapprox 0.5,
then the bulge is classical) and a few are less reliable (star formation rate cannot be used for S0s).  But,
by and large, it is reasonable to conclude that the use of $M$ criteria, each with failure probability~$\epsilon_m$, 
results in a classification with a failure probability of order the product of the individual
failure probabilities, $\Pi_1^M \epsilon_m$.  This becomes very small very quickly as $M$ grows even
to 2 and especially to $M > 2$.  For example, essentially all bulge-pseudobulge classifications in
KH13 were made using at least two and sometimes as many as five criteria.

      Fisher \& Drory (2015) also contribute new criteria that become practical as new technology 
such as intergral-field spectroscopy gets applied to large samples of galaxies.  These are incorporated 
into an enlarged list of classification criteria below.

      A shortcoming of Fisher \& Drory's approach is that it is applied without regard to galaxy Hubble types.  
But we know that both many S0s and many Sbcs contain pseudobulges, but the latter all tend to be star-forming 
whereas the former generally are not.  This is one reason for their conclusion (e.{\ts}g.)~that high star
formation rate near the galaxy center robustly implies a pseudobulge, but no star formation near the center 
fails to prove that the bulge is classical.  Classification criteria that involve gas content and star formation 
rate cannot be applied to S0 galaxies.  Application to Sas is also fragile.  Fortunately, most criteria do work
for early-type galaxies.

\subsection{Enlarged List of Bulge-Pseudobulge Classification Criteria}
\label{subsec:2}


      Kormendy \& Kennicutt (2004), Kormendy (2012), and Fisher \& Drory (2015) together provide the following 
improved list of (pseudo)bulge classification criteria.  I note again: The failure rate for individual criteria 
ranges from 0\ts\% to roughly 25\ts\%.  Therefore the use of more criteria quickly gives much more reliable results.

\begin{enumerate}

\item[(1){\kern -3pt}]{If the galaxy center is dominated by young stars and gas
                       but there is no sign~of a merger in progress, then the bulge is mostly pseudo. 
                       Ubiquitous star formation must be secular.  Fisher \& Drory (2015) make this
                       quantitative: if the specific star formation rate sSFR $\geq$ $10^{-11}$ yr$^{-1}$,
                       then the bulge is likely to be pseudo; whereas if sSFR $<$ $10^{-11}$ yr$^{-1}$, then
                       the bulge is likely to be classical.  Also, if the bulge is very blue,
                       $B - V < 0.5$, then it is pseudo.  Criteria (1) cannot be used for S0s. }

\item[(2){\kern -3pt}]{Disky pseudobulges (a) generally have apparent flattening similar to that of the outer disk 
                       or (b) contain spiral structure all the way to the galaxy center.  Classical bulges 
                       are much rounder than their disks unless they are seen almost face-on, and they cannot 
                       have spiral structure.  Criterion 2(a) can be used for S0s; 2(b) can not.}

\item[(3){\kern -3pt}]{Pseudobulges are more rotation-dominated than are classical bulges 
                       in the \hbox{$V_{\rm max}/\sigma$\ts--\ts$\epsilon$} diagram; $V_{\rm max}$ is
                       maximum rotation velocity, $\sigma$ is near-central velocity dispersion, and
                       $\epsilon$ is ellipticity.
                       Integral-field spectroscopy often shows that the central surface brightness excess over the inward 
                       extrapolation of the disk profile is a flat central component that rotates rapidly and has 
                       small $\sigma$.}

\item[(4){\kern -3pt}]{Many pseudobulges are low-$\sigma$ outliers in the Faber-Jackson (1976) correlation between (pseudo)bulge 
                       luminosity and velocity dispersion.  Integral-field~spectra often show that $\sigma$ 
                       decreases from the disk into a pseudobulge.  Fisher and Drory make this 
                       quantitative: Pseudobulges have rather flat logarithmic derivatives of the dispersion
                       profile $d{\log{\sigma}}/d{\log{r}} \geq -0.1$ and $V^2/\sigma^2 \geq 0.35$.  
                       In contrast, if $d{\log{\sigma}}/d{\log{r}} < -0.1$ or if central 
                       $\sigma_0 > 130$ km s$^{-1}$, then the bulge is classical.

\item[(5){\kern -3pt}]{Small bulge-to-total luminosity ratios do not guarantee that a bulge
                       is pseudo,~but almost all pseudobulges have $PB/T$ \lapprox \ts0.35.~If $B/T$ \gapprox \ts0.5, 
                       the bulge is classical.}

\item[(6){\kern -3pt}]{Most pseudobulges have S\'ersic index $n < 2$; most classical bulges have $n \geq 2$.}

\item[(7){\kern -3pt}]{Classical bulges fit the fundamental plane correlations for elliptical
                       galaxies.  Some pseudobulges do, too, and then the correlations are not useful for
                       classification.  More extreme pseudobulges are fluffier than classical bulges;
                       they have larger effective radii $r_e$ and fainter effective surface brightnesses $\mu_e$.
                       These pseudobulges can be identified using fundamental plane correlations.}

\item[(8){\kern -3pt}]{In face-on galaxies, the presence of a nuclear bar shows that a pseudobulge dominates the central light.  
                       Bars are disk phenomena.  Triaxiality in giant Es involves different physics -- 
                       slow (not rapid) rotation and box (not $x_1$ tube) orbits.  }

\item[(9){\kern -3pt}]{In edge-on galaxies, boxy bulges are edge-on bars; seeing one identifies a pseudobulge.  
                       The boxy-core-nonrotating side of the ``E{\ts}--{\ts}E dichotomy'' between two kinds of elliptical 
                       galaxies (see Section 4.1.1) cannot be confused with boxy, edge-on bars 
                       because boxy ellipticals -- even if they occur in disk galaxies (we do not know of an example) 
                       -- are so luminous that we would measure $B/T > 0.5$.  Then point (5) would tell us that this 
                       bulge is classical.}

\item[(10){\kern -3pt}]{Fisher \& Drory (2015) conclude that pseudobulges have weak Fe and Mg b lines:
                       equivalent width of [Fe $\lambda$5150\ts\AA]\ts$<$\ts3.95\ts\AA; equivalent width of [Mg{\ts}b]\ts$<$\ts2.35\ts\AA.
                       In their sample, no classical bulge has such weak lines.  Some pseudobulges have 
                       stronger lines, so this criterion, like most others, is not 100\ts\% reliable.}

\item[(11){\kern -3pt}]{If a bulge deviates from the [Mg b]\ts--\ts$\sigma$ or [Mg b]\ts--\ts[Fe] correlations 
                       for elliptical galaxies by $\Delta$[Mg b] $< 0.7$ -- that is, if the [Mg] line strength is
                       lower than the scatter for Es -- then the bulge is likely to be pseudo (Fisher \& Drory 2015).}}

\end{enumerate}

      It is important to emphasize that classical and pseudo bulges can occur together.  Fisher \& Drory (2015) review examples 
of dominant pseudobulges that have small central classical bulges.  And some giant classical bulges contain nuclear disks (e.{\ts}g., 
NGC 3115: Kormendy \etal 1996b; NGC 4594: Kormendy \etal 1996a).

      Criterion (9) for boxy pseudobulges works only for edge-on and near-edge-on galaxies.
In face-on galaxies, it is easy to identify the elongated parts of bars, but they also have rounder, denser
central parts, and these are not easily distinguished from classical bulges (Athanassoula 2015; Laurikainen
\& Salo 2015).  So the above criteria almost certainly fail to find some pseudobulges in face-on barred galaxies.

\vskip -30pt

\centerline{\phantom{00000000000000000}}

\subsection{Secular Evolution in Disk Galaxies: Applications}

\vskip -15pt

\centerline{\phantom{00000000000000000}}

      Progress in many subjects depends on a full integration of the picture of disk secular evolution into 
our paradigm of galaxy evolution.  Examples include the following:

\begin{enumerate} 

\item[(1){\kern -3pt}]{If the smallest bulges are pseudo and not classical, then the luminosity and mass functions of
                       classical bulges and ellipticals are very bounded: 
                       \hbox{$M_K$ \lapprox $-19$;} \hbox{$M_V$ \lapprox $-16$;}
                       $L_V$ \gapprox $10^{8.5}$ $L_\odot$; stellar mass $M_{\rm bulge}$ \gapprox $10^9$ $M_\odot$.
                       In simulations (Brooks \& Christensen 2015; Section 4 here),
                       the physics that makes classical bulges and ellipticals does not need to explain objects that
                       are smaller than the above.  More accurately: If the same generic physics (e.{\ts}g.,
                       major mergers) is relevant for smaller objects, it does not have to produce remnants that
                       are consistent with low-mass extrapolations of parameter correlations for classical bulges
                       and ellipticals.  One possible reason may be that the progenitors of that physics are very gas-rich.
                       \lineskip=-15pt \lineskiplimit=-15pt}

\item[(2){\kern -3pt}]{Our understanding that, below the above limits, lower-mass bulges are essentially all pseudo
                       makes it harder to understand how galaxy formation by hierarchical clustering of CDM 
                       makes so many giant, classical-bulge-less (i.{\ts}e., pure-disk) galaxies.  This was
                       the theme of the observational papers Kormendy \etal (2010) and Fisher \& Drory (2011).  It is
                       addressed in Brooks \& Christensen (2015).  We return to this issue in Section 4.}

\item[(3){\kern -3pt}]{Understanding how supermassive black holes (BHs) affect galaxy evolution
                       requires an understanding that classical and pseudo bulges are different.  
                       Classical bulges participate in the correlations between BH mass and 
                       bulge luminosity, stellar mass, and velocity dispersion.  Pseudobulges essentially do not.
                       This is some of the evidence that BHs coevolve with classical bulges and
                       ellipticals in ways to be determined, whereas BHs exist in but do not influence
                       the evolution of disks or of disk-grown pseudobulges.  We return to this subject in 
                       Section 6.}

\end{enumerate}

\vfill\eject

\section{Giant Clumps in High-z Gas-Rich Disks Make Classical Bulges}

      The second major advance in our picture of bulge formation involves the observation that many high-$z$ disks 
are very gas-rich and dominated by \hbox{$10^8$\ts--\ts$10^9$ $M_\odot$, kpc-size} star-forming clumps 
(Elmegreen{\ts}et{\ts}al.\ts2005,\ts2007,\ts2009a,{\ts}b; 
Bournaud{\ts}et{\ts}al.\ts2007;
Genzel{\ts}et{\ts}al.\ts2006,\ts2008,\ts2011;
F\"orster{\ts}Schreiber{\ts}et{\ts}al.\ts2009,\ts2011a,{\ts}b;
Tacconi{\ts}et{\ts}al. 2010).
These galaxies evidently accrete cold gas so rapidly that they become violently unstable.  Bulgeless disks 
tend to have small epicyclic frequencies $\kappa$.  If the surface density $\Sigma$ rapidly grows
large and is dominated by gas with low velocity dispersion~$\sigma$, then the  Toomre (1964) 
instability parameter $Q = 0.30 \sigma \kappa/ G \Sigma$ \lapprox \ts1 ($G$ = graviational constant).  
The observed clumps are interpreted to be the result.  Theory and simulations suggest that the 
clumps sink rapidly toward the center by dynamical friction.  
They also dump large amounts of additional cold gas toward the center via tidal torques.  The result 
is violent relaxation plus a starburst that produces a classical bulge.  Many papers discuss this evolution (e.{\ts}g.,
Dekel, Sari, \& Ceverino 2009;
Ceverino, Dekel, \& Bournaud 2010;
Cacciato, Dekel, \& Genel 2012;
Forbes \etal 2014;
Ceverino \etal 2015).
Bournaud (2015) reviews this subject in the present book.  I include it here for two reasons,
it is a major advance, so it deserves emphasis in this concluding chapter, and I want to add two science points:

      Figure 1 illustrates my first point:~{\it Evolution by clump sinking, inward gas transport, violent relaxation, 
and starbursts proceeds much as it does in our picture of wet major mergers.  That is, in practice (if not in its beginnings),
classical bulge formation from clump instabilities is a variant of our standard picture of bulge formation in wet major mergers.}
The process starts differently than galaxy mergers~-- \phantom{000000000000000}

\centerline{\null} \vfill  

\begin{figure}[hb]



  \includegraphics{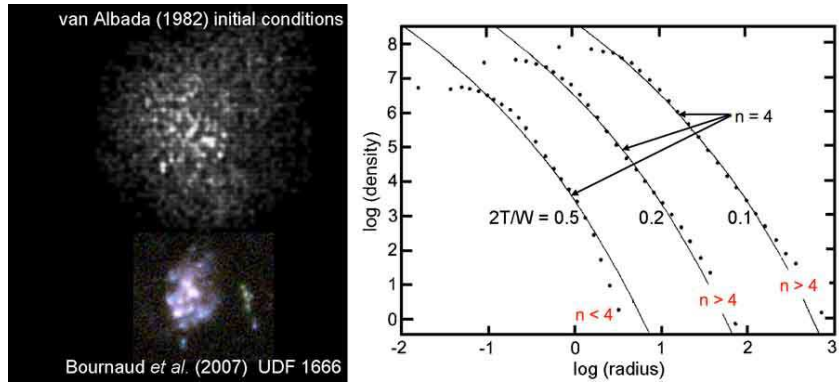}

\caption{Mergers of clumpy initial conditions make S\'ersic (1968)
         function remnants with indices $n \sim 2${\ts}--{\ts}4. A remarkably early illustration is 
         the $n$-body simulation of van Albada (1982), whose initial conditions (grayscale densities)
         resemble the clumpy high-$z$ galaxy UDF 1666 studied by Bournaud \etal (2007).  
         Van Albada's initial conditions were parameterized by the ratio
         of twice the total kinetic energy~to the negative of the potential energy.~In 
         equilibrium, $2T/W$\ts$=$\ts1.  For smaller values,
         gentle collapses ($2T/W = 0.5$) make S\'ersic profiles with~$n < 4$.
         Violent collapses ($2T/W$\ts\lapprox\ts0.2) make $n$\ts\gapprox 4.  Clump sinking
         in high-$z$ disks is inherently gentle.  The hint is that the clumps merge to make classical bulges
         with $n < 4$.  This figure is from Kormendy (2012).
         }

\end{figure}

\eject

\noindent what merges here are not finished galaxies but rather are clumps that formed quickly and temporarily in unstable disks.
          Nevertheless, what follows -- although two- and not three-dimensional -- is otherwise closely similar to 
          a wet merger with gas inflow and a starburst.  That is, it is a slower, gentler version of Arp 220.

          Early models by Elmegreen \etal (2008) confirm that gas-rich galaxy disks violently form 
clumps like those observed.  The clumps quickly sink, merge, and make a high-S\'ersic-index, vertically thick bulge.  
It rotates slowly, and rotation velocities decrease with increasing distance above and below the disk plane.
These are properties of classical bulges, and Elmegreen and collaborators conclude that this process indeed
makes classical (not pseudo) bulges.  Many of the later papers summarized above and reviewed by Bournaud (2015)
reach similar conclusions.

      However, Bournaud (2015) goes on to review more recent simulations that\ts-- among other improvements -- 
include strong feeback from young stars.  The results complicate the above picture.  For example,
Genel \etal (2012) find that ``galactic winds are critical for [clump] evolution.  The giant clumps we obtain are
short-lived and are disrupted by wind-driven mass loss.  They do not virialize or migrate to the galaxy centers
as suggested in recent work neglecting strong winds.''  Other simulations produce pseudobulge-like, small S\'ersic
indices.  Some results are inherently robust, such as the conclusion that gas-rich, violently unstable disks
at high $z$ gradually evolve into gas-poor, secularly evolving disks at lower redshifts 
(Cacciato, Dekel, \& Genel 2012;
cf.~Ceverino, Dekel, \& Bournaud 2010).
However, the conclusions from the models are substantially more uncertain than the inferences from the observations.
This is part of a problem that I emphasize in the next section: 

      Simulations of baryonic galaxy evolution inside CDM halos formed via $n$-body simulations of cosmological 
hierarchical clustering are making rapid progress as the baryonic physics gets implemented in better detail.
But these simulations still show clearcut signs of missing important physics.  In contrast, practitioners of
this art who carefully put great effort into improving the physics tend to be overconfident about its results.
We are -- I will suggest -- still in a situation where robust observational conclusions that are theoretically
squishy are more trustworthy than conclusions based on state-of-the-art simulations, at least when baryonic
physics is involved.

      Another caveat is the observation that the clumps in high-$z$ disks are much less
obvious in the inferred mass distributions than they are in rest-frame optical or blue light (Wuyts \etal 2012).
Frontier observations have opened up a popular new window on the formation of classical bulges,
but its importance is not entirely clear.

      In the present subject of bulge formation, it seems provisionally plausible that
formation via high-$z$ disk instabilites and consequent clump sinking represents a significant new channel
in the formation of classical bulges.  Meanwhile, a large body of work from the 1980s and 1990s continues
to tell us that major galaxy mergers make classical bulges, too.  Can we distinguish the results of the
two processes?  We do not yet know, but my second point is that Figure 1 provides a hint: Although results
are still vulnerable to unknown details in (for example) feedback, it seems likely that the classical
bulges produced by sinking clumps have S\'ersic indices that are systematically smaller than those made 
by major galaxy mergers.  This is one aspect of many that deserves further work.~See also point (8) in Section\ts8.

\vfill\eject

\section{Making Classical Bulges and Ellipticals by Major Mergers}

      Brooks \& Christensen (2015) is perhaps the most important chapter in~this~book.  The mainstream 
of theoretical work on galaxy formation has come to be the simulation in a cosmological context first of 
purely collisionless CDM but now with gloriously messy baryonic physics included.  Progress is impressively rapid, 
but we are far from finished.  This subject is well reviewed from the perspective of its practitioners by 
Brooks and Christensen.  This includes a discussion of uncertainties and shortcomings in the models, 
again as seen by theorists.  As an observer, I have a complementary perspective on which 
measurements of galaxies provide the most useful constraints on and ``targets'' for formation models. 
It gives me the feeling that modelers are at least partly ``barking up the wrong tree.''
This section complements Brooks and Christensen (2015) by reviewing these observations.

      Pseudobulge formation was covered in Section 2.  Here, I focus on the formation of classical bulges
and ellipticals.  My discussion uses the observations that classical bulges are essentially indistinguishable 
from coreless-disky-rotating ellipticals (see, e.{\ts}g., Figure 4).  The inference is that they
formed in closely related ways.  


\vskip -25pt

\centerline{\phantom{00000000000000000}}

\subsection {Observer's Perspective on Bulge Formation Via Major Mergers}

      I begin with giant ellipticals and classical bulges:~their structure and formation 
are understood in the most detail.  Classical bulges are identified by the criteria listed in 
Kormendy \& Kennicutt (2004),
Kormendy (2012),
KH13,
Fisher \& Drory (2015), and
Section 2 here.  
I know no observational reason to seriously doubt our understanding of bulges with $B/T$ \gapprox 0.8.
Then, as $B/T$ drops to \lapprox \ts1/2, the situation gets less clear.  Our formation 
picture may still essentially be correct, but it gets less directly based on observations as $B/T$ or 
bulge luminosity decreases.  Meanwhile, the theoretical problem is that simulations make too many bulges,
especially big ones.  In this section, I review things that we know and outline things that we do not know.  
It is critically important to start with a discussion of ellipticals, because
our understanding of classical bulges must be within this context.

\vskip -25pt

\centerline{\phantom{00000000000000000}}

\subsubsection {Observed Properties of Ellipticals: Clues to Their Formation}

      The observed properties of elliptical galaxies are reproduced by simulations of wet and dry mergers 
in remarkable detail.  These are not embedded in large-scale cosmological simulations, but this is 
not a fundamental fault if the initial conditions are realistic -- galaxies with typical $z \sim 0$ 
gas fractions and encounter velocities that are roughly parabolic.  
Kormendy \etal (2009, hereafter KFCB) provide an ARA\&A-style review and develop some of the evidence.  
Hopkins \etal (2009a and 2009b) provide the most detailed models for wet and dry mergers, respectively.
These papers are comprehensive; a concise summary of the ``E -- E dichotomy'' in Kormendy (2009) is updated below.
The critical observation is that ellipticals come in two varieties and that bulges are similar to one (but not both)
of these varieties.

      The E\ts--{\ts}E dichotomy of ellipticals into two kinds is based on these observations: \vskip 5pt

      \underbar{Giant ellipticals} ($M_V$ \lapprox \ts$-21.5 \pm 1$ for $H_0 = 70$ km s$^{-1}$ Mpc$^{-1}$)
generally \hfill\break
(1) have S\'ersic function outer profiles with $n > 4$; \hfill\break
(2) have cores; i.{\ts}e., central missing light with respect to the outer S\'ersic profile;\hfill\break
(3) rotate slowly, so rotation is of little importance dynamically; hence \hfill\break
(4) are anisotropic and modestly triaxial; \hfill\break
(5) are less flattened (ellipticity $\epsilon$ $\sim${\thinspace}0.2) than smaller ellipticals; \hfill\break
(6) have boxy-distorted isophotes; \hfill\break
(7) mostly are made of very old stars that are enhanced in $\alpha$ elements (Figure 2); \hfill\break
(8) often contain strong radio sources (Figure 3), and \hfill\break
(9) contain X-ray-emitting gas, more of it in more luminous galaxies (Figure 3). \vskip 5pt

      \underbar{Normal ellipticals and dwarf ellipticals like M{\ts}32} ($M_V$ \gapprox \ts$-21.5$) generally \hfill\break
(1) have S\'ersic function outer profiles with $n \simeq 2$ to 3; \hfill\break
(2) are coreless -- have central extra light with respect to the outer S\'ersic profile;\hfill\break
(3) rotate rapidly, so rotation is dynamically important to their structure;  \hfill\break
(4) are nearly isotropic and oblate spheroidal, albeit with small axial dispersions; \hfill\break
(5) are flatter than giant ellipticals (ellipticity $\epsilon$ $\sim${\thinspace}0.35); \hfill\break
(6) have disky-distorted isophotes; \hfill\break
(7) are made of younger stars with little $\alpha$-element enhancement (Figure 2); \hfill\break
(8) rarely contain strong radio sources (Figure 3), and \hfill\break
(9) generally do not contain X-ray-emitting gas (Figure 3). \vskip 5pt

      These results are etablished in many papers (e.{\thinspace}g.,
Davies et al.~1983;
Bender 1988;
Bender et al.~1989; 
Nieto et al.~1991;
Kormendy et al.~1994; 
Lauer et al.~1995, 2005, 2007a, b;
Kormendy \& Bender 1996;
Tremblay \& Merritt 1996;
Gebhardt et al.~1996; 
Faber et~al.~1997; 
Rest et al.~2001;
Ravindranath et al.~2001;
Thomas et al.~2002a, b, 2005;
Emsellem et al.~2007, 2011;
Cappellari et al.~2007, 2011, 2013b;
KFCB;
Kuntschner \etal 2010).
A few ellipticals are exceptions to one or more of (1) -- (9).  The above summary is quoted from Kormendy (2009).

     Why is this relevant here?  The answer is that 
classical bulges are closely similar to coreless-disky-rotating ellipticals.  No bulge is similar to a 
core-boxy-nonrotating elliptical as far as I know.  This is a clue to formation processes.
First, though, we need to understand the difference between the two kinds of ellipticals:

     How did the E{\thinspace}--{\thinspace}E dichotomy arise?  The ``smoking gun'' for an explanation
is a new aspect of the dichotomy originally found in Kormendy (1999) and observed in all low-luminosity ellipticals
in the Virgo cluster by KFCB.  Coreless galaxies do not have featureless power-law profiles.  Rather, 
all coreless galaxies in the KFCB sample show a new structural component, i.{\thinspace}e., central extra light 
above the inward extrapolation of the outer S\'ersic profile.  Kormendy (1999) suggested that the
extra light is produced by starbursts fed by gas dumped inward during dissipative mergers.  Starbursts were predicted by merger 
simulations as soon as these included gas, dissipational gas inflow, and star formation (Mihos \& Hernquist 1994).  
Mihos and Hernquist were concerned that extra components had not been observed.  The reason turns 
out to be that we had not measured ellipticals with enough surface brightness range and spatial resolution.~Like
Faber{\ts}et{\ts}al.\ts(1997,\ts2007), KFCB suggest that the origin of the E\ts--{\ts}E dichotomy is
that core ellipticals formed in dry mergers whereas coreless ellipticals formed in wet mergers.  Simulations of 
dry and wet mergers reproduce the structural properties of core and extra light ellipticals in beautiful 
detail (Hopkins et al.~2009a, b).  And, although the formations scenarios differ, Khochfar \etal (2011) similarly
conclude that the difference between fast and slow rotators is related to cold gas dissipation and star-formations shutdown, respectively.

      Cores are thought to be scoured by supermassive black hole binaries that
were formed in major mergers.  The orbit shrinks as the binary flings stars away.  This decreases the surface 
brightness and excavates a core
(Begelman \etal 1980; 
Ebisuzaki \etal 1991; 
Makino \& Ebisuzaki 1996; 
Quinlan \& Hernquist 1997; 
Faber \etal 1997; 
Milosavljevi\'c \& Merritt 2001; 
Milosavljevi\'c et al.~2002; 
Merritt 2006).  The same process should happen during wet mergers; in fact, gas accelerates the orbital decay 
(Ivanov, Papaloizou, \& Polnarev 1999;
Gould \& Rix 2000;
Armitage \& Natarajan 2002, 2005;
Escala \etal 2004, 2005;
Dotti \etal 2007;
Hayasaki 2009;
Cuadra \etal 2009; 
Escala \& Del Valle 2011; see
Mayer 2013 for a recent review).  
However, we observe that the fraction
of the luminosity that is in extra light in low-luminosity ellipticals is larger than the fraction of the light
that is ``missing'' in the cores of high-luminosity ellipticals.  KFCB suggest that core scouring is
swamped by the starburst that makes the extra light in coreless-disky-rotating ellipticals.

      When did the E{\thinspace}--{\thinspace}E dichotomy arise?  Figure 2 shows observation (7) that core ellipticals 
mostly are made of old stars that are enhanced in $\alpha$ elements.  In contrast, coreless ellipticals 
are made of younger stars with more nearly solar compositions.  This means (Thomas et al.~2002a, b, 2005) that the stars 
in core Es formed in the first few billion years of the universe and over a period of \lapprox\ts1 Gyr,
so quickly that Type I supernovae did not have time to dilute with Fe the 
\hbox{$\alpha$-enriched} gas recycled  \phantom{000000000000}


\vfill


  \includegraphics{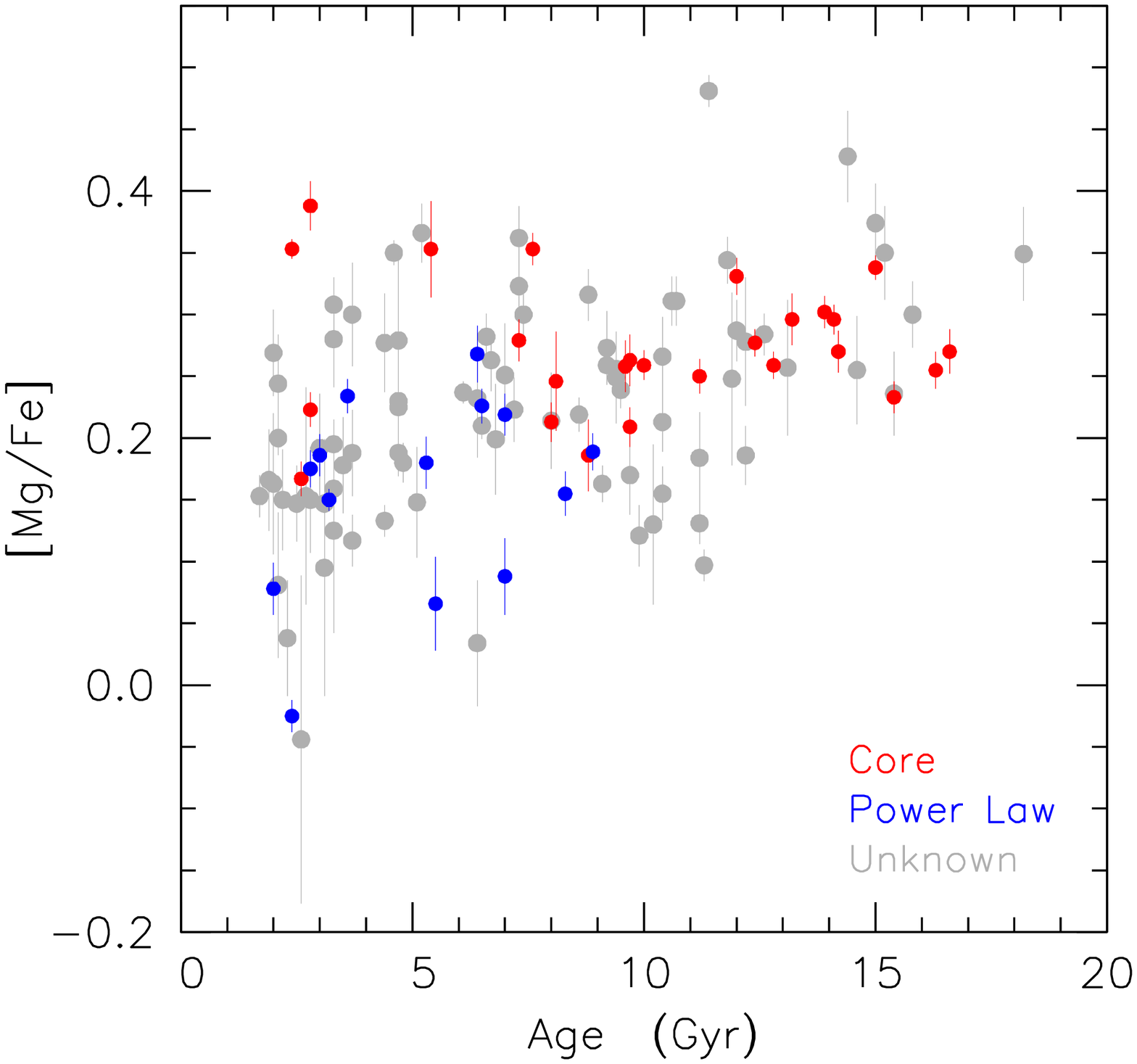}

\begin{figure}
\caption{
 \pretolerance=15000  \tolerance=15000 
 \lineskip=0pt \lineskiplimit=0pt
Alpha element overabundance ($\log$ solar units) versus relative age of the stellar
population.  Red and blue points denote core and ``power law'' (i.{\ts}e., coreless) ellipticals. 
The [Mg/Fe] and age data are from Thomas \etal (2005); this figure is from KFCB.
}\end{figure}

\eject

\noindent by Type II supernovae.  This does not mean that core ellipticals were made at the same time as their stars.
Mass assembly via dry mergers as required to explain their structure could have happened at 
any time after star formation stopped.  Our problem is to explain how star formation was 
quenched so quickly and not allowed to recur.  In contrast, coreless ellipticals have younger, 
less-$\alpha$-enhanced stellar populations.  They are consistent with a simple picture in 
which a series of wet mergers with accompanying starbursts formed their stellar populations 
and assembled the galaxies more-or-less simultaneously over the past 9 billion years.  
Faber et al.~(2007) discuss these issues in detail.  A big problem with the present state of the art
is that we know so little about mergers and merger progenitors at high $z$.  

          Why did the E{\thinspace}--{\thinspace}E dichotomy arise?~The key observations are:~(8) 
core-boxy ellipticals often are radio-loud whereas coreless-disky ellipticals are not, and 
(9) core-boxy ellipticals contain \hbox{X-ray} gas whereas coreless-disky ellipticals do not
(Bender et al.~1989).  Figure 3 (from KH13) illustrates these results.  KFCB suggest 
that the hot gas keeps dry mergers dry and protects giant ellipticals from late star formation.  This
is the operational solution to the above ``maintenance problem''.   I~return to the problem of
star-formation quenching in Section 7.

\vfill




 \includegraphics{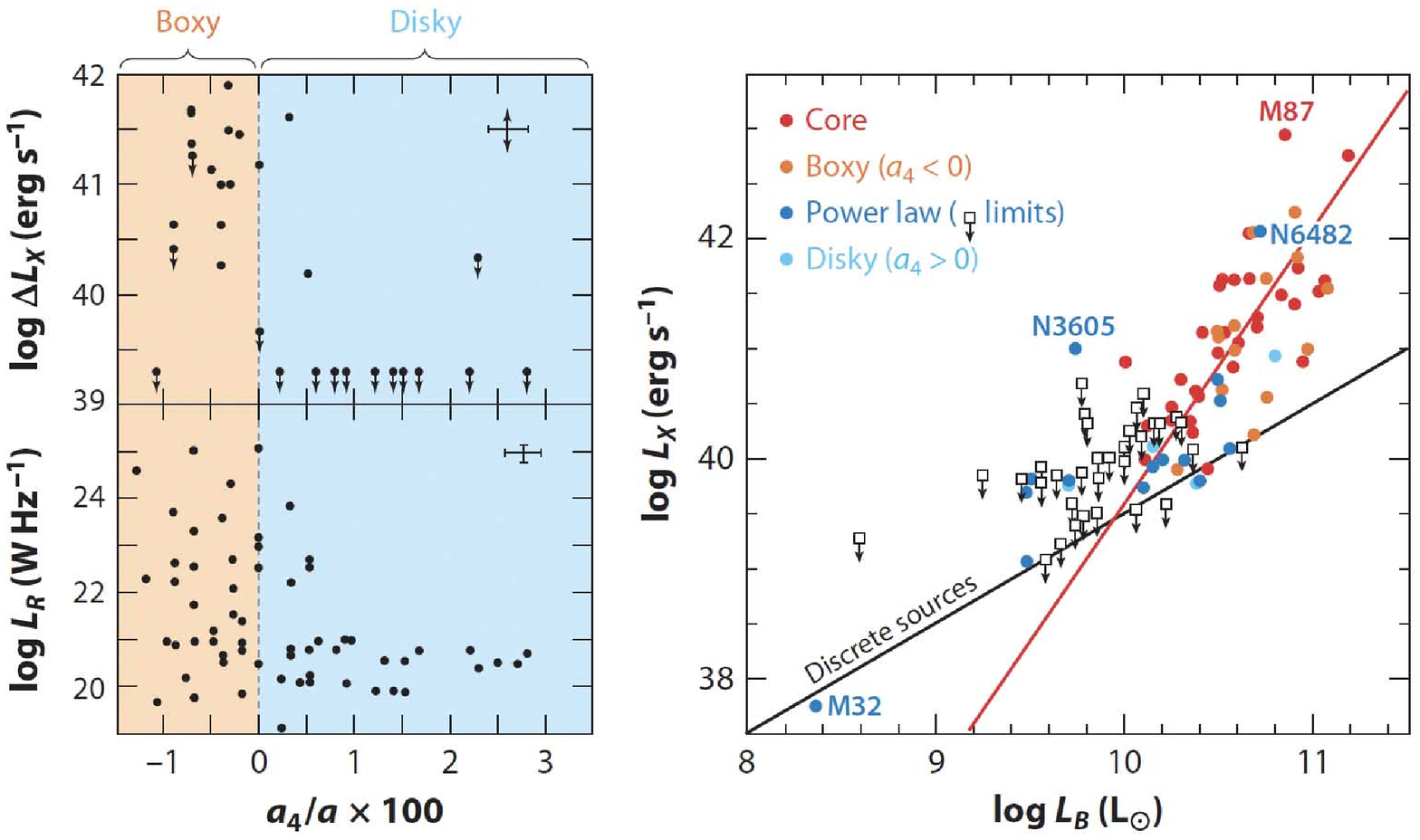}

\begin{figure}
\caption{
 \pretolerance=15000  \tolerance=15000 
 \lineskip=0pt \lineskiplimit=0pt
(Left) Correlation with isophote shape parameter $a_4$ of (top) X-ray emission from hot gas and
(bottom) radio emission (from Bender \etal 1989).  Boxy ellipticals ($a_4 < 0$) contain hot gas and
strong radio sources; disky ellipticals ($a_4 > 0$) generally do not.  
(Right) KFCB update of the X-ray correlation.  Detections are color-coded according to the E\ts--{\ts}E 
dichotomy.  The emission from X-ray binary stars is estimated by the black line (O'Sullivan, Forbes,
\& Ponman 2001); this was subtracted from the total emission in constructing the left panels.
The red line is a bisector fit to the core-boxy-nonrotating ellipticals.  They statistically
reach $L_X = 0$ from hot gas at $\log{L_B} \simeq 9.4$.  This corresponds to $M_V \simeq -20.4$,
a factor of 2 fainter than the luminosity that divides the two kinds of ellipticals.  Thus,
if a typical core E was made in a merger of two equal-mass galaxies, then both were marginally
big enough to contain X-ray gas and the remnant immediately was massive enough so that hot gas
could quench star formation.  KFCB suggest that this is why these mergers were dry.
For similar results, see Pellegrini (1999, 2005) and Ellis \& O'Sullivan (2006).
}\end{figure}

\eject

     In the above story, the challenge is to keep the hot gas hot, given that X-ray gas cooling times are short (Fabian 1994).  
KFCB review evidence that the main heating mechanism may be energy feedback from accreting BHs (the active galactic
nuclei [AGNs] of observation 8); these may also have helped to quench star formation.  Many details of this picture require work 
(Cattaneo et al.~2009).  Cosmological gas infall is an additional heating mechanism 
(Dekel \& Birnboim 2006).  Still, Figure 3 is a crucial connection 
between \hbox{X-ray} gas, AGN physics, and the E\ts--{\ts}E dichotomy.

      ``\underbar{Bottom line}:'' In essence, only giant, core ellipticals and their progenitors 
are massive enough to contain hot gas that helps to engineer the E{\thinspace}--{\thinspace}E dichotomy. 

\vskip 12pt

\begin{svgraybox}
\phantom{0000000000}
\end{svgraybox}

\vskip -68pt
\phantom{0000000000}

\subsubsection{Classical Bulges Resemble Coreless-Disky-Rotating Ellipticals}

\vskip -30pt
\phantom{0000000000}
\begin{svgraybox}
\phantom{0000000000}
\end{svgraybox}

\vskip -65pt
\phantom{0000000000}

\begin{svgraybox}

      Are both kinds of ellipticals also found as bulges?  So far, observations indicate that the answer is ``no''.
Classical bulges closely resemble only the coreless-disky-rotating ellipticals.  There are apparent exceptions in the
literature, but all the exceptions that I know about are classification errors brought about (e.{\ts}g.)~by the very 
large S\'ersic indices of some core galaxies (see KFCB Table 1 for examples and KFCB Section 5.2 for discussion).
This comment also does not include ellipticals with nuclear disks.  All signs are that these involve different physics,
so these really are ellipticals, not S0 bulges.
 
      There is physics in this conclusion.  The X-ray gas prevents cooling and dissipation during any subsequent 
mergers or any $z$\ts\lapprox 1 cold accretion.  Plausibly, it should also prevent there from being any cold gas left over to make 
a new disk after a merger is complete.  Further checks, both of the observational conclusion and of the
theoretical inference, should be made.

\end{svgraybox}

\vskip -20pt

\begin{svgraybox}
\phantom{0000000000}
\end{svgraybox}

\vskip -45pt
\phantom{0000000000}

\subsubsection{The Critically Important Target for Galaxy Formation}

\vskip -30pt
\phantom{0000000000}

\begin{svgraybox}

      The most fundamental distinction between galaxy types is the one between bulges $+$ ellipticals and disks.
Bulges and disks overlap over a factor of about $\sim$\ts1500 in luminosity and mass (Figure 4), but over that entire 
overlap range, they are dramatically different from each other. This includes differences in specific angular momentum 
(Romanowsky \& Fall 2012; Fall \& Romanowsky 2013), in orbit structure, in flattening, and in radial density profiles (disks are roughly 
exponential; coreless-disky-rotating ellipticals have $n$ \gapprox \ts2).  At absolute magnitude $M_V \simeq -16.7$ and outer
circular-orbit rotation velocity $V_{\rm circ} \sim 85$ km s$^{-1}$, M{\ts}32 is a normal small elliptical
galaxy (KFCB).  At $M_V \simeq -21.6$ and $V_{\rm circ} = 210 \pm 15$ km s$^{-1}$, M{\ts}101 is almost 100 times 
more luminous but is thoroughly different from M{\ts}32 (Kormendy \etal 2010). 

\end{svgraybox}

\begin{svgraybox}
      I believe that the goal of galaxy formation modeling should be to produce realistic disks and realistic ellipticals
that overlap over the observed factor of $\sim$\ts1500 in luminosity but that differ as we observe them to differ over the 
whole of that range.  And over the whole of that range, disks and bulges can be combined with $B/T$ and $D/T \simeq 1 - B/T$
ratios that have the observed distribution (i.{\ts}e., $B/T \sim 1$ near the upper end of the range, but it can be $\ll$\ts1
at the bottom of the range).  The properties of individual disks and bulges are essentially independent of $B/T$ with
structural parameters shown in Figure\ts4.

\end{svgraybox} 

\vfill

\includegraphics{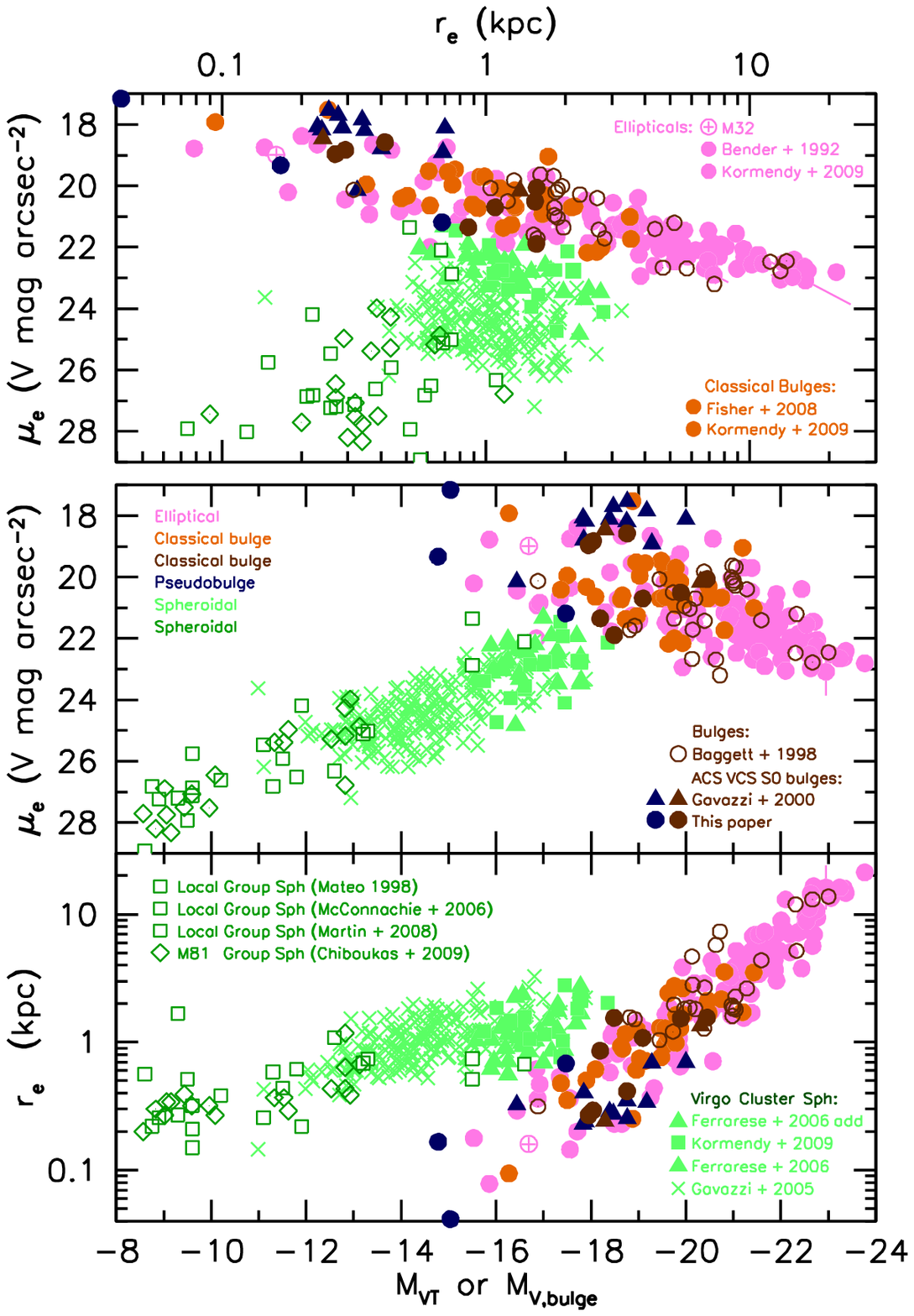}

\includegraphics{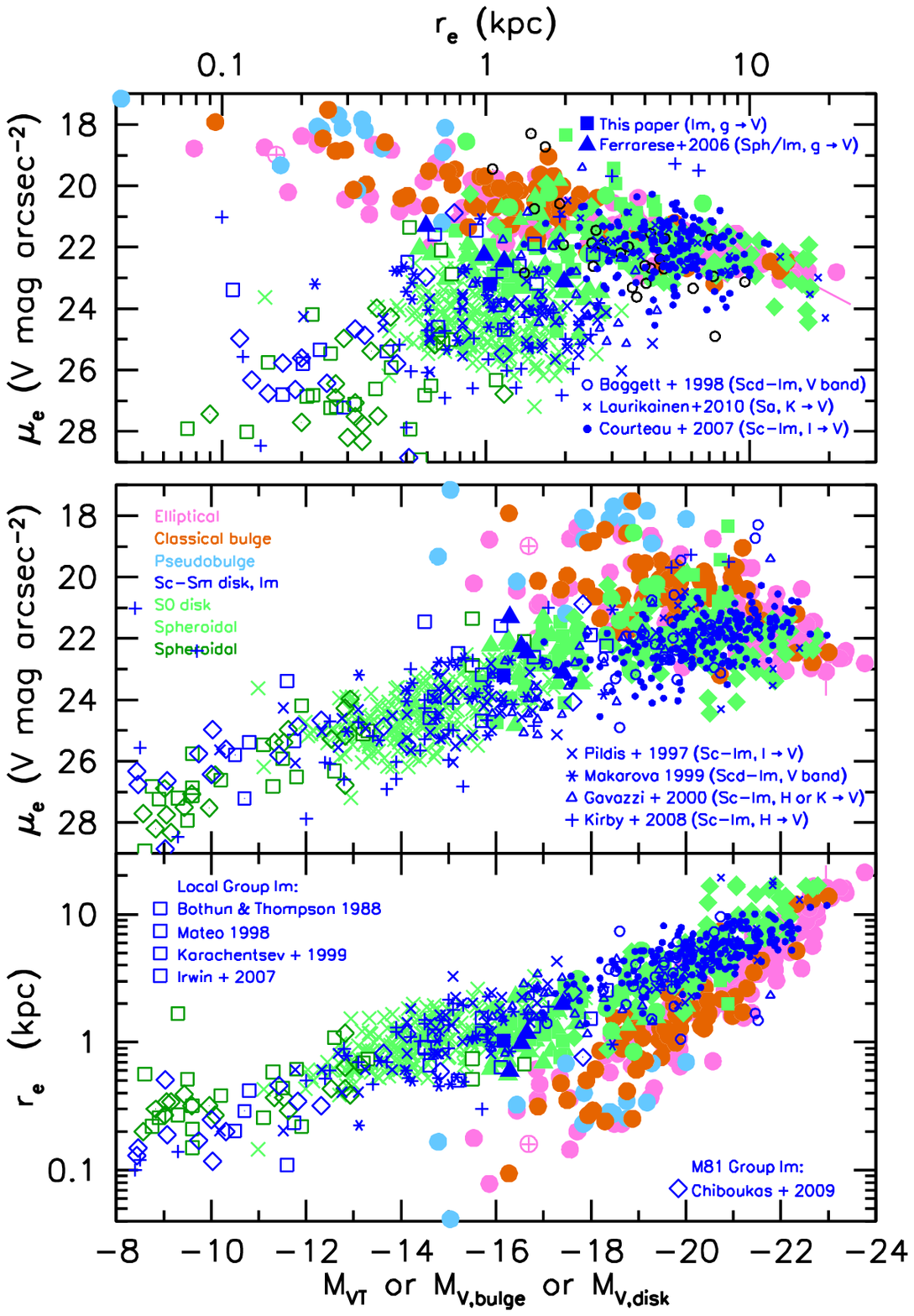}

\begin{figure}
\caption{
 \pretolerance=15000  \tolerance=15000 
 \lineskip=0pt \lineskiplimit=0pt
Correlations between effective radius $r_e$, effective brightness $\mu_V$, 
and absolute magnitude $M_V$ for classical bulges and ellipticals (brown and pink points), for
spheroidal (Sph) galaxies and S0 disks (green points), and for spiral galaxy disks (blue points).  
When bulge-disk decomposition is necessary, the two components are plotted separately.  Bulges and
disks overlap from $M_V \simeq -15$ to $M_V \simeq -23$, i.{\ts}e., over a factor of about 1500.  The left
panel shows (1) that Sph galaxies are distinct from bulges $+$ ellipticals and (2) that classical bulges
and ellipticals satisfy the same structural parameter correlations.  The right panel adds S0 and S
galaxy disks.  It shows that all disks satisfy the same structural parameter correlations
over the whole range of luminosities.  Note that disks and ellipticals have similar $r_e$ and $\mu_e$ 
at the highest luminosities, but they have very different S\'ersic indices ($\sim 1$ and 2 to $>$\ts10,
respectively).  As a result, the central surface brightnesses in bulges and ellipticals are
more than an order of magnitude higher than the central surface brightnesses of disks (Kormendy
1985, 1987).  Bulges $+$ ellipticals and disks also have non-overlapping distributions of intrinsic flattening
(e.{\ts}g., Sandage, Freeman, \& Stokes 1970).  From Kormendy \& Bender (2012),
}
\end{figure}

\eject

      Of course, bulges and disks are not different in {\it every} parameter; e.{\ts}g., the $r_e$\ts--\ts$\mu_e$
correlations overlap at high luminosities (Figure 4).  This makes sense: At the highest masses, it does not 
require much dissipation to turn a disk into an elliptical, at least in terms of virial parameters.  All that
is required is to scramble disk orbits into an ellipsoidal remnant.  A larger amount of dissipation is 
required to make the high-density centers, and in {\it central} parameters and parameter correlations, disks and 
bulges $+$ ellipticals are very different (Kormendy 1985, 1987).

\vskip -10pt

\begin{svgraybox}
\phantom{0000000000}
\end{svgraybox}

\vskip -69pt
\phantom{0000000000}

\subsubsection{Critical Obserational Clue: The Problem of Giant, Pure-Disk Galaxies Depends on Environment,
               Not on Galaxy Mass}

\vskip -34pt
\phantom{0000000000}

\begin{svgraybox}

      The most difficult challenge in our picture of galaxy formation\ts--{\ts}I suggest\ts--{\ts}is to understand
how hierarchical clustering produces so many giant, pure-disk galaxies that have no sign of 
a classical bulge.  CDM halos grow by merging; fragments arrive from all directions, and not all
fragments are small.  There are two parts to this problem:  (1) It is difficult to understand how
cold, flat disks survive the violence inherent in the mergers that grow DM halos.  And
(2) it is difficult to prevent the stars that arrive with the latest accretion victim from adding 
to a classical bulge that formed in (1) from the scrambled-up disk.

      Brooks \& Christensen (2015) review how the modeling community tries to solve this problem.
In spite of several decades of evidence that mergers make bulges, they do not use mergers to turn
disks into bulges.  Instead, they use feedback from young stars and active galactic nuclei to ``whittle
away'' the low-angular-momentum part of the distribution of gas angular momenta and argue that this
prevents bulge formation.  And they use feedback to delay disk formation until the halo is assembled.  
Feedback is likely to be important in the formation of dwarf galaxies (Governato \etal 2010), and indeed, 
they essentially never have bulges (e.{\ts}g., Kormendy \& Freeman 2015, Figure 10).  

      However, it is difficult for me to believe that feedback, either from star formation or from AGNs, is 
responsible for the difference between bulges and disks.  Feedback is fundamentally an internal process 
that is controlled by the galaxy's potential well depth.  It is not clear how {\it only} tweaking the 
feedback can make a small elliptical like M{\ts}32 (different from small disks) and a giant disk like M{\ts}101
(different from similarly giant ellipticals) with no intermediate cases.  Bulge-to-total ratios vary widely,
but classical bulges are always like ellipticals no matter what the B/T ratio, and disks are always different 
from ellipticals no matter what the D/T ratio.  Observations do not suggest that it is primarily feedback 
that results in this difference.  Rather:

      There is a fundamental observational clue that modelers are not using:

      Whether evolution makes disks or whether it makes bulges does not depend mainly on galaxy mass.
Rather, it is a strong function of environment.  Kormendy \etal (2010) show that, in the extreme field 
(i.{\ts}e., in environments like the Local Group), most giant galaxies ($V_{\rm circ} \geq 150$ km s$^{-1}$) 
are pure disks.  Only 2 of 19 giant galaxies closer to us than 8 Mpc have $B/T$ as big as 1/3.  Only 2 more 
are ellipticals.  A few have smaller classical bulges, but 11 of the 19 galaxies have essentially no classical 
bulge.  In contrast, $>$ 2/3 of all stars in the Virgo cluster live in bulges or elliptical galaxies. 
{\it There is no problem of understanding giant pure-disk galaxies in the Virgo cluster.}  It is a
mature, dense environment that contains large amounts of X-ray-emitting, hot gas.  Rich clusters are places 
where most of the baryons live suspended in hot gas  (e.{\ts}g., Kravtsov \& Borgani 2012).  I argue in Section 4.1.1
that various heating processes maintain this situation for very long times.  In contrast, poor groups are
environments in which accretion of cold gas from the cosmic web can dominate, as long as the galaxies involved -- 
i.{\ts}e., the aforementioned pure disks -- are low enough in mass so that they cannot hold onto X-ray gas.

      As long as this environmental dependence is not a primary, essential part of the explanation, I believe 
that attempts to solve the problem of overproduction of bulges in $\Lambda$CDM cosmology are ``barking up the wrong tree''.

\end{svgraybox}

\quad Why can't we use feedback to delay star formation until the halo is assembled?  As reviewed by
Brooks \& Christensen (2015), this is commonly suggested.  The counterexample is our Galaxy: The oldest stars in 
the thin disk are $\sim$\ts$10^{10}$ yr old, so much of the growth of our Galaxy happened when the thin disk was 
already in place (Kormendy \etal 2010, p.~73).  
      
\subsubsection{It is not a problem that major mergers are rare}

      The prevailing theoretical paradigm is more and more converging on the view that major mergers are rare -- are, 
in fact, almost irrelevant -- and that, instead, minor mergers make both bulges and ellipticals (see Naab 2013 for a review),
even some core-boxy-nonrotating ellipticals (Naab \etal 2014).  It will be clear from this writeup that, based on observational 
evidence, I agree that major mergers are rare.  But I disagree that they are unimportant in the formation of bulges and ellipticals.

      The above papers make important points that are robust.  They argue convincingly that major mergers are rare -- that
only a small fraction of galaxies undergo several of them in their recent history (say, since $z \sim 2$).  And many
authors argue that most star formation does not occur during mergers; rather, it occurs in a ``main sequence'' 
of disks of various masses, with higher star formation rates at higher masses (e.{\ts}g.,
Schiminovich \etal 2007;
Noeske \etal 2007;
Elbaz \etal 2007;
Daddi \etal 2007;
Finlator \& Dav\'e 2008;
Karim \etal 2011;
Peng \etal 2010;~Rodighiero~et~al.~2011;
Wuyts \etal 2011;
Salmi \etal 2012;
Whitaker \etal 2012;
Tacconi \etal 2013;
Speagle \etal 2014).
These authors conclude that the duty cycle of star formation is large.  Therefore most star formation does not occur
in rare events.  I made the same argument in Section 2: If almost all galaxies of a particular type are energetically 
forming stars, then star formation must be secular; it cannot be episodic with short duty cycles.  Caveat: the star 
formation that is associated with mergers is not instantaneous. Puech \etal (2014) argue that merger-induced star
formation is significant.  Are these results consistent with a picture in which essentially all formation of classical bulges 
and ellipticals happens via major mergers? 

      I believe that the answer is yes, although the details need further work.  Elliptical galaxies are observed to be rare; 
the morphology-density relation (Dressler 1980; Cappellari \etal 2011) shows that they are a small fraction of all galaxies
except in rich clusters.  Classical bulges are rarer than we thought, too; this is a clear conclusion of the work on disk 
secular evolution.  {\it Therefore the events that make bulges must be rare.}  It is also not a problem if most 
star formation happens~in~disks.  For example, only a small fraction of the galaxy mass is contained in the extra light 
components that are identified by KFCB and by Hopkins (2009a) as the parts of coreless/disky/rotating ellipticals that 
formed in the most recent ULIRG-like starburst (e.{\ts}g., Genzel \etal 2001).  Most of the mass was already in stars before these late, 
wet mergers.  And in dry mergers, essentially all the mass was already in stars (or in X-ray gas that stays X-ray gas) and essentially 
no new stars are formed. 

      How many mergers do we need to explain elliptical galaxies?  Toomre (1977) already pointed out that a reasonable
increase in merger rate with increasing $z$ would suffice.  He based this on ten mergers-in-progress that he
discussed in his paper.  He assumed that such objects are identifiable for $\sim$ half a billion years.  Then, if the number
of mergers in progress increased as (lookback time)$^{5/3}$ consistent~with a flat distribution of binding energies
for galaxy pairs, the result is that the number of remnants is consistent with the number of elliptical and early-type
disk galaxies.  This estimate was made for the level of completeness of the Second Reference Catalogue of Bright Galaxies
(de Vaucouleurs, de Vaucouleurs, \& Corwin 1976).  

      Conselice (2014) reviews observational estimates of how merger rates depend on $z$.  As Toomre predicted,
the major merger rate is inferred -- e.{\ts}g., from counting close pairs of galaxies -- to increase rapidly with $z$.
Observations of high-$z$ galaxies show that close binary fractions increase roughly as $(1 + z)^m$ with $m \sim 2$ to 3 (e.{\ts}g.,
Bluck \etal 2009, 2012;
Conselice \etal 2009;
L\'opez-Sanjuan \etal 2013;
Tasca \etal 2014).
ULIRGs increase in comoving energy density even faster toward higher redshift, at least out to $z = 1$ (Le Floc'h \etal 2005).
The necessary connections between these results to establish or disprove whether bulges $+$ ellipticals are made via major 
mergers have not been established.  Important uncertainties include (1) the low-mass end of the mass functions for ellipticals
and especially for classical bulges, and (2) the degree to which mass-clump sinking in disks contributes.  However, the above
results on merger frequencies appear at least qualitatively consistent with the conclusion that bulges and ellipticals are
made in major mergers, as the pre-2000 history of observational work established (see Schweizer 1998 for a review).

      A shortcoming of many current investigations is that they concentrate on a few parameter distributions for large galaxy samples 
and not specifically on the histories of bulges and disks.  E.{\ts}g., they look at the statistics of what fraction of galaxies experience mergers.  
Outcomes are difficult to estimate, because with samples of $10^4$ to $10^5$ $z \sim 0$ galaxies or $10^2$ high-$z$ galaxies, the typical galaxy
is only a few pixels in radius.  Then it is difficult to identify and classify galaxy components.

\subsubsection{Uncertainties With Our Picture of Bulge Formation in Major Mergers}

\centerline{\phantom{00000000000000000}}

\vskip -18pt

      Two major uncertainites are a concern (see also Brooks \& Christensen 2015).
Virtually all observational evidence on mergers-in-progress (e.{\ts}g.,
Toomre 1977;
Joseph \& Wright 1985;
Sanders \etal 1988a, b; 
Hibbard \etal 1994, 1995, 1996, 2001a, b; see 
Schweizer 1987, 1990, 1998 for reviews) 
involves giant galaxies.  And the detailed evidence is for $z \sim 0$ galaxies with gas
fractions of a few to $\sim$\ts10\ts\%.  (1) We do not have comparable evidence for dwarfs.  That is, we have 
not studied a sample of dwarfs that fill out a merger sequence from close pairs to mergers engaged 
in violent relaxation to train wrecks that are still settling down to mature objects.  And (2) we do not have comparably
detailed studies of galaxies at high z that have gas fractions \gapprox \ts50\%.  It is possible that mergers behave 
differently for such objects.

\vskip -26pt

\centerline{\phantom{00000000000000000}}

\subsubsection{The Problem of Giant, Pure-Disk Galaxies: Conclusion}

\centerline{\phantom{00000000000000000}}

\vskip -18pt

      My most important suggestion in this section is that the modeling community relies too strongly on feedback 
as the only way to prune excessive bulge formation.  On the contrary, I suggest that environmental differences in 
the amount of dynamical violence in galaxy formation histories are the central factor.  I suggest that the solution 
is not a to whittle away the low-angular-momentum tail of the distribution of angular momenta in forming galaxies.  
Nearby galaxies dramatically show us the importance of violent relaxation.  To me, the issue is: How much does
violent relaxation dominate? How much is the evolution controlled by gentle accretion?  And how do the answers
depend on environment?

\vskip -34pt

\centerline{\phantom{00000000000000000}}

\section{Universal Scaling Relations For All Galaxies?}

\centerline{\phantom{00000000000000000}}

\vskip -20pt

      How we best construct parameter correlations depends on what~we~want~to~learn.  Projections~of the 
fundamental plane correlations {\it separate} galaxy classes; e.{\ts}g., bulges$+$ellipticals from disks$+$Sphs (Figure 4). 
So they teach us about differences in formation processes.  In contrast, it is possible to construct parameter 
correlations that make most or all galaxy types look continuous.  {\it These encode less information about galaxy formation}.
E.{\ts}g., in a projection of the structural parameter correlations that encodes mass-to-light ratio, the difference
between ellipticals, spheroidals, and even irregulars largely disappears (Bender, Burstein, \& Faber 1992).  Zaritsky (2015) 
regards this as progress -- as replacing correlations that are flawed with ones that capture some inherent simplicity.  
That simplicity is real.  But it is insensitive to the power that other correlations clearly have to tell us things 
about galaxy formation.

      I therefore disagree, not with Zaritsky's operational results but with his motives.  If you look at the
fundamental plane face-on, it contains lots of information.  If you look at it edge-on, then it looks simple.  
This may feel like a discovery.  But it just means that you are looking at a projection that hides the information content 
in the parameter plane.  Other combinations of parameters make still more types of objects looks continuous and indistinguishable.  
But this means that we learn still less, not more, about their nature and origin.  The simple correlations are not uninteresting,
but the ones that teach us the most are the ones that correctly identify differences that turn out to have causes within
formation physics.

\section{Coevolution of Supermassive Black Holes and Host Galaxies}

      The observed demographics of supermassive black holes (BHs) and their implications for the coevolution (or not)
of BHs and host galaxies are discussed in Kormendy \& Ho (2013).  This is a 143-page ARA\&A review that revisits methods 
used to measure BH masses $M_\bullet$ using spatially resolved stellar and gas dynamics.  It also provides a detailed 
analysis of host galaxy morphologies and properties.  Careful treatment of the $M_\bullet$ and galaxy measurements 
allows Kormendy and Ho to reach a number of new science conclusions.  They are summarized in this section.

      Graham (2015) reviews the same subject in the present book.  Some of his review is historical,
especially up to the beginning of his Section 4.1 but also sporadically thereafter.  I do not comment here on 
the historical review.  However, on the science, I cannot ``duck'' my responsibility as author of this concluding chapter{\kern 0.3pt}:

      I disagree with most  of the scientific conclusions in Graham (2015).  Starting in his Section 4.1,
his discussion uses data and repeats conclusions from Graham \& Scott (2013, 2015).  Problems with the 2013 data 
are listed in KH13 (p.\ts555); a point made there that is not repeated further here is that 
many of Graham's galaxy classifications are incorrect.  Here, rather than write a point-by-point rebuttal to 
Graham (2015), I first concentrate on a summary of the unique strengths of the KH13 analysis and data.  
However, a few comments are added to further explain the origin of the disagreements with Graham (2015).
I then summarize the KH13 results and conclusions about $M_\bullet$\ts--{\ts}host-galaxy correlations
(Sections 6.1 and 6.2).

\begin{svgraybox}

\centerline{\null} \vskip -20pt

      Before I begin, a comment is in order about how readers react to disagreements in the literature.
The most common reaction is that the subject needs more work.  Specialists may know enough to decide
who is correct.  But the clientele community of non-specialists who mainly want to use the results 
often do not delve into the details deeply enough to decide who is correct.  Rather, their reaction is 
that this subject needs further work until everybody agrees that the disagreement is resolved.  Sometimes, 
this is an appropriate reaction, when the issues are more complicated than our understanding of the physics, 
or when measurements are still too difficult, or when results under debate have low significance compared to
statistical errors or systematic effects.  My reading of the community is that reactions to disagreements 
on BH demographics take this form.

      However, I suggest that we already know enough to decide who is correct in the disagreement
between KH13 and Graham (2015).  Our ARA\&A review and the Graham \& Scott papers both
provide enough detail to judge the data and the analysis.  It is particularly important to note how these
separate discussions do or do not connect up with a wide body of results in other published work, including
other chapters in this book.  A strength of the Kormendy \& Ho analysis is that it connects up with -- i.{\ts}e.,
it uses and it has implications for -- a wide variety of aspects of galaxy formation.

\end{svgraybox}

      Strengths of the data and suppoprting science that are used by KH13 include the following. 
Some of these points are discussed more fully in the Supplemental Material of KH13. 

\begin{enumerate} 

\item[(1){\kern -3pt}]{BH masses based on absorption-line spectroscopy are now derived by including
                       halo dark matter in the stellar dynamical models.  This generally leads to an upward 
                       revision in $M_\bullet$ by a factor that can be \gapprox \ts2 for core galaxies.
                       Kormendy and Ho use these masses.  For some galaxies (e.{\ts}g., M{\ts}87), Graham 
                       uses them; for other galaxies (e.{\ts}g., NGC 821, NGC 3377, NGC 3608, NGC 4291,
                       NGC 5845), he does not, even though such masses are published (Schulze \& Gebhardt 2011).
                       \lineskip=-15pt \lineskiplimit=-15pt}

\item[(2){\kern -3pt}]{Kormendy and Ho include new $M_\bullet$ determinations for mostly high-mass galaxies from
                       Rusli \etal (2013).  Graham \& Scott (2013) did not include these galaxies.  It is not 
                       clear whether they are included in Graham (2015), but observation that the highest $M_\bullet$
                       values plotted in his Figure 4 are $\sim 6 \times 10^9$ $M_\odot$ and not $>$\ts$10^{10}$\ts$M_\odot$
                       suggests that they are not included, at least in this figure.
                       \lineskip=-15pt \lineskiplimit=-15pt}

\item[(3){\kern -3pt}]{BH masses derived from emission-line gas rotation curves are used without correction
                       when the emission lines are narrow.  However, when the emission lines are
                       wide -- often as wide in km s$^{-1}$ as the rotation curve amplitude -- some authors
                       have ignored the line widths in the $M_\bullet$ determinations.  KH13 argue
                       that these BH masses are underestimated and do not use them.  Graham (2015) uses them. 
                       \lineskip=-15pt \lineskiplimit=-15pt}

\item[(4){\kern -3pt}]{All disk-galaxy hosts have $B/T$ values based on at least one and sometimes as
                       many as six bulge-disk decompositions.  Graham \& Scott (2013) use a mean statistical correction 
                       to derive some bulge magnitudes from total magnitudes.
                       \lineskip=-15pt \lineskiplimit=-15pt}

\item[(5a){\kern -3pt}]{All disk-galaxy hosts have (pseudo)bulge classifications that are based on at least two and
                        as many as five criteria such as those listed here in Section 2.1.  Graham (2015) rejects this
                        approach and instead compares BH--host correlations for barred and unbarred galaxies.  However,
                        Kormendy and Ho emphasize that some barred galaxies contain classical bulges, whereas many
                        unbarred galaxies contain pseudobulges.  If classical and pseudo bulges correlate differently
                        with their BHs (Figure 7), then a division into barred and unbarred galaxies does not cleanly see this.
                        It should be noted that other derivations of BH--host correlations (e.{\ts}g., the
                        otherwise very good paper by McConnell \& Ma 2013), also do not differentiate between classical and
                        pseudo bulges.  They compare early and late galaxy types.  But many S0s contain pseudobulges,
                        and a few Sbcs contain classical bulges (e.{\ts}g., NGC 4258: Kormendy \etal 2010).}

\item[(5b){\kern -3pt}]{\pretolerance=100000\tolerance=100000 The picture of disk secular evolution and the conclusion that 
                           \hbox{pseudobulges}
                           are distinguishable from classical bulges is fully 
                           integrated into the analysis.  Graham (2011, 2015) does not use this picture and
                           argues that classical and pseudo bulges cannot reliably be
                           distinguished.  Kormendy \& Kennicutt (2004), Kormendy (2012), Kormendy \& Ho (2013),
                           Fisher \& Drory (2015) in this book, and Section 2 in this summary chapter
                           disagree.  \hbox{The subject is growing} rapidly, and whole meetings are devoted to it
                           (e.{\ts}g., 2012 IAU General \hbox{Assembly} Special Session 3, ``Galaxy Evolution 
                           Through Secular Processes,'' {\tt http://bama.ua.edu/$\sim$rbuta/iau-2012-sps3/proceedings.html} and
                           Kormendy 2015;
                           XXIII Canary Islands Winter School, ``Secular Evolution of Galaxies'', Falc\'on-Barroso \&
                           Knapen 2012).  Kormendy \& Ho make a point of distinguishing classical and pseudo bulges by purely morphological
                           criteria such as those given in Section\ts2.1.  The fact that we then discover that
                           BHs correlate differently with classical and pseudo bulges is a substantial 
                           success of the secular evolution picture.
                       \lineskip=-15pt \lineskiplimit=-15pt}

\item[(6){\kern -3pt}]{KH13 find that the $\log M_\bullet$ -- $M_{K,\rm bulge}$, 
                       $\log M_\bullet$ -- $\log \sigma$ and $\log M_\bullet$\ts--\ts$\log M_{\rm bulge}$ correlations
                       for classical bulges and ellipticals have intrinsic scatter of 0.30, 0.29, and 0.28 dex,
                       respectively.  This small scatter is a consequence of the care taken in (1)\ts--{\ts}(5), above, 
                       in implementing a uniform, accurate distance scale based as much as possible on standard candles,
                       in correcting galaxy classifications when detailed photometry reveals errors, 
                       and in correcting $K$-band magnitudes for systematic errors.
                       Given this small scatter, it was possible to discover a new result; i.{\ts}e., that five sample
                       galaxies that are major mergers in progress deviate from the above correlations in having undermassive
                       BHs for their host size (see Figure 14 in KH13).  Having noted this result, the five
                       mergers are omitted from our correlation fits shown below.  However, mergers in progress are included in 
                       Graham (2015) and in McConnell \& Ma (2013).  
                       \lineskip=-15pt \lineskiplimit=-15pt}

\end{enumerate}

      These procedural differences plus others summarized in KH13 or omitted here for the sake
of brevity account for most of the differences in the correlation plots shown in Graham (2015) and those in KH13.
Generically, they have the following effects (ones in italics also apply to McConnell \& Ma 2013).
(1) {\it At the high-$M_\bullet$ end, Graham's BH masses are biased low, because he uses underestimated values from 
emission-line rotation curves,} because he uses $M_\bullet$ values that are not corrected for effects of halo dark matter,
and because he does not consistently use the Rusli \etal (2013) high-$M_\bullet$ galaxies.  (2) At the \hbox{low-$M_\bullet$}
end, Graham's BH masses are biased low, because he includes pseudobulges.  Differentiating barred and unbarred galaxies
is not sufficient to solve this problem.  {\it McConnell and Ma also include pseudobulges, differentiating early- and
late-type galaxies helps, although many S0s contain pseudobulges.}  (3) Graham regards M{\ts}32 as pathological and 
omits~it.  KFCB show that it is a normal, tiny elliptical.  Including it in KH13 helps to anchor the
BH correlations at low BH masses.  (4) {\it The result is that the BH--host correlations have much larger scatter in Graham (2015)
and in McConnell \& Ma (2013) than they do in KH13 (see Figures 5 and 7 below)}.  Also, Graham sees a
kink in the \hbox{$\log M_\bullet$ -- $M_{K,\rm bulge}$} correlation whereas we do not, and he sees no kink in the 
$\log M_\bullet$ -- $\log \sigma$ whereas we see signs of a kink at high $\sigma$ where $M_\bullet$ becomes largely
independent of $\sigma$.  McConnell \& Ma (2013) and KH13 agree on the kinks (and lack of kinks) in the
$M_\bullet$\ts--{\ts}host-galaxy correlations.

\subsection{Correlations Between BH Mass and Host Galaxy Properties from Kormendy \& Ho (2013)}

      This section summarizes the BH{\ts}--{\ts}host-galaxy correlations from KH13.

\vfill\eject

      The procedures summarized above lead in KH13 to Table 2 for 44 elliptical galaxies and 
Table 3 for 20 classical bulges and 21 pseudobulges.  Figure 5 shows the resulting $\log M_\bullet$ -- $M_{K,\rm bulge}$ and
$\log M_\bullet$ -- $\log \sigma$ correlations for classical bulges and ellipticals.  Mergers in progress are omitted as
explained~above, and three ``monster'' BHs that deviate above the correlations are illustrated in faint symbols but
are omitted from the fits.  Also shown are symmetric, least-squares fits (Tremaine \etal 2002) symmetrized
around $L_{K,\rm bulge}$\ts=\ts$10^{11}$\ts$L_{K\odot}$ and $\sigma_e$\ts=\ts200{\ts}km{\ts}s$^{-1}$:
\vskip -28pt
\null
$$
\log\biggl(\kern -2pt{{M_\bullet} \over {10^9{\ts}M_\odot}}\kern -2pt\biggr) 
              = -(0.265 \pm 0.050) - (0.488 \pm 0.033) (M_{K,\rm bulge} + 24.21);  \eqno{(1)}
$$
\centerline{\null} 
\vskip -42pt 
\centerline{\null}
$$
\log\biggl(\kern -2pt{{M_\bullet} \over {10^9{\ts}M_\odot}}\kern -2pt\biggr) 
              = -(0.509 \pm 0.049) + (4.384 \pm 0.287) \log \biggl({{\sigma} \over {200~{\rm km~s}^{-1}}}\biggr). \eqno{(2)}
$$                   
\null
\vskip -14pt
\noindent Here, we adopt equal errors of $\Delta M_{K,\rm bulge} = 0.2$ and $\Delta \log{M_\bullet} = 0.117$, i.{\ts}e., the mean 
for all fitted galaxies.  Then the intrinsic scatters in Equations (1) and (2) are 0.30 dex and 0.29 dex, respectively.  In
physically more transparent terms,

\vskip -5pt
$$
{{M_\bullet} \over {10^9~M_\odot}} = \biggl(0.544^{+0.067}_{-0.059}\biggr)\ 
                                           \biggl({{L_{K,\rm bulge}} \over 
                                                  {10^{11}{\ts}L_{K\odot}}}\biggr)^{1.22 \pm 0.08} \eqno{(3)}
$$
\null
\vskip -26pt
\null
$$
{{M_\bullet} \over {10^9~M_\odot}} = \biggl(0.310^{+0.037}_{-0.033}\biggr)\ 
                                           \biggl({{\sigma} \over 
                                                  {200~{\rm km~s}^{-1}}}\biggr)^{4.38 \pm 0.29} \eqno{(4)}
$$

\noindent Both relations have shifted to higher BH masses because of corrections to $M_\bullet$, because mergers in 
progress are omitted, and because pseudobulges are postponed.

\vskip 10pt

\vfill


 \includegraphics{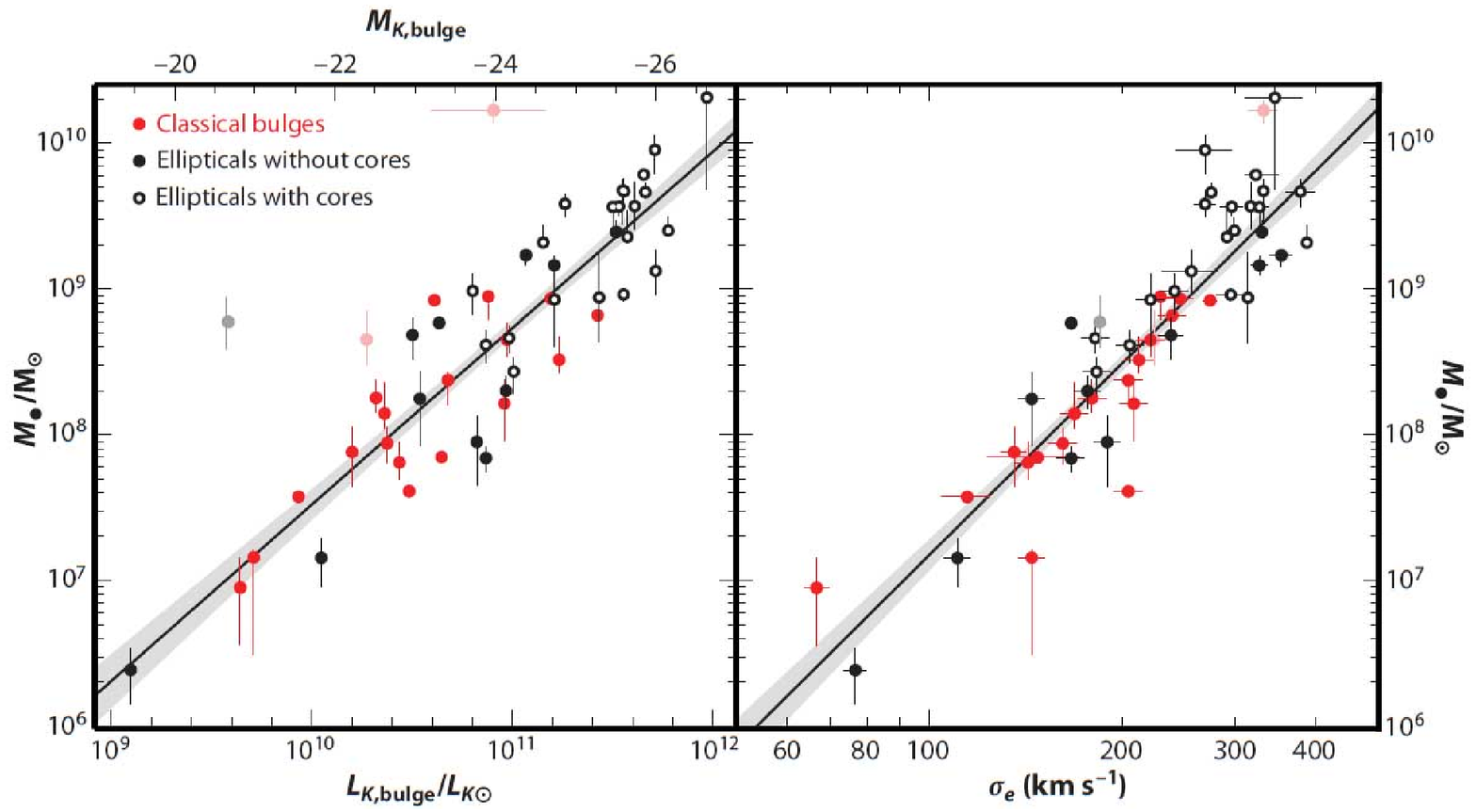}

\begin{figure}
\caption{
 \pretolerance=15000  \tolerance=15000 
 \lineskip=0pt \lineskiplimit=0pt
Correlations of BH mass $M_\bullet$ with the $K$-band absolute magnitude and luminosity of the host bulge ({\it left panel}) 
and with its velocity dispersion at radii where $\sigma_e$ is unaffected by the BH ({\it right panel}).  Black points are 
for ellipticals; a white center indicates that this galaxy has a core.  Red points are for classical bulges.  The lines  
are Equations~(1)~and~(2).  Note:~the $M_\bullet$--$M_{K,\rm bulge}$ correlation remains log-linear with no kink 
at high luminosities.  In contrast, the biggest BH masses look essentially independent of $\sigma_e$ in ellipticals that have cores.
From KH13.
}
\end{figure}

\eject

\centerline{\null}

      The $\log M_\bullet$ -- $L_{K,\rm bulge}$ correlation in Figure 5 is converted to a correlation with bulge stellar
mass $M_{\rm bulge}$ by applying mass-to-light ratios that were engineered by KH13 to be independent
of the papers that determine $M_\bullet$, to have zeropoints based on the Williams \etal (2009) dynamical models,
but also to take variations in stellar population age into account.  The resulting mass correlation is:
\vskip -18pt
\null
$$
\null\quad\quad100\biggl({{M_\bullet} \over {M_{\rm bulge}}}\biggr) = \biggl(0.49^{+0.06}_{-0.05}\biggr)\ 
                                           \biggl({{M_{\rm bulge}} \over 
                                                  {10^{11}{\ts}M_\odot}}\biggr)^{0.15 \pm 0.07},~~ \eqno{(5)}
$$
\null
\vskip -6pt
\noindent with an intrinsic scatter of 0.28 dex.  The BH mass fraction, 
$M_\bullet/M_{\rm bulge}$\ts=\ts$0.49^{+0.06}_{-0.05}$\ts\% at $M_{\rm bulge} = 10^{11}$\ts$M_\odot$, 
is approximately a factor of 4 larger than we thought before the $M_\bullet$ values were corrected 
(Merritt \& Ferrarese 2001; 
Kormendy \& Gebhardt 2001;
McClure \& Dunlop 2002;
Marconi \& Hunt 2003;
Sani \etal 2011).

Note again that $M_\bullet$--$L_{\rm bulge}$ is a single power law with no kink, whereas $M_\bullet$--$\sigma$ 
is a power law that ``saturates'' at high $M_\bullet$ (see also McConnell \& Ma 2013).  That is, $M_\bullet$ 
becomes nearly independent of $\sigma$ in the highest-$\sigma$ galaxies that also have cores (Figure 5).
We understand why: The Faber-Jackson $L$--$\sigma$ correlation saturates at high $L$, because $\sigma$
does not grow very much once galaxies are massive enough so that all mergers are dry (Figure 6).
This is seen in simulations of dry, major mergers by (e.{\ts}g.)~Boylan-Kolchin, Ma, \& Quataert (2006)
and by Hilz \etal (2012).  Section 4.1.1 reviewed arguments why core ellipticals are remnants of dry mergers.

\vfill



 \includegraphics{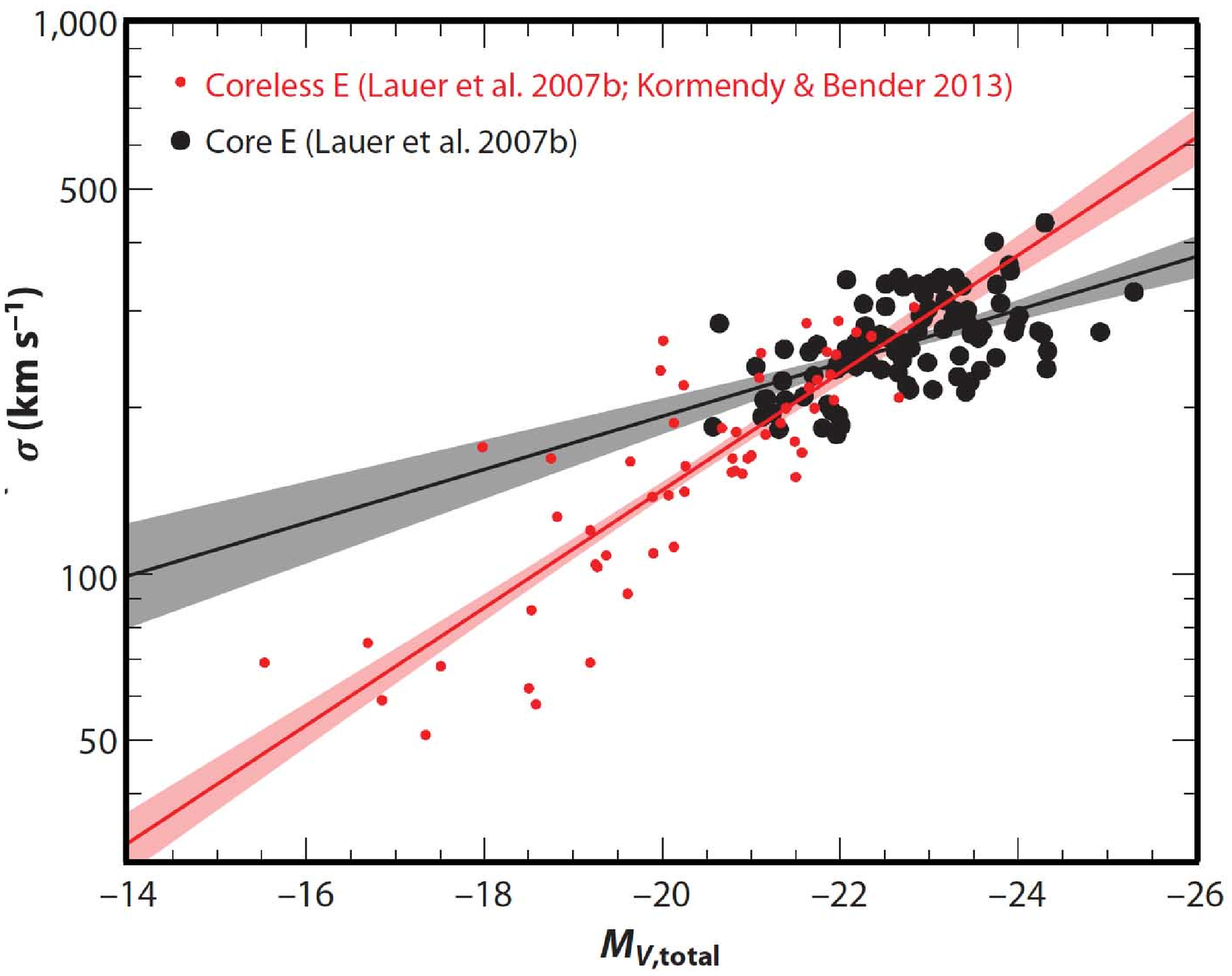}

\begin{figure}
\caption{
Faber-Jackson (1976) correlations for core ellipticals ({\it black\/}) and coreless ellipticals ({\it red\ts}).  Total $V$-band 
absolute magnitudes $M_{V,\rm total}$, velocity dispersions $\sigma$, and profile types are mostly from Lauer \etal (2007b) or 
otherwise from KFCB.  The lines are symmetric least-squares fits to core Es ({\it black line}) and coreless Es ({\it red line}) 
with 1-$\sigma$ uncertainties shaded.  The coreless galaxies show the familiar relation, $\sigma \propto L_V^{0.27 \pm 0.02}$.  
But velocity dispersions in core ellipticals increase only very slowly with luminosity, $\sigma \propto L_V^{0.12 \pm 0.02}$.  
As a result, $M_\bullet$ becomes almost independent of $\sigma$ for the highest-$\sigma$ galaxies in Figure 5.
This figure from KH13 is based on Kormendy \& Bender (2013).  Lauer \etal (2007a) and Cappellari \etal
(2013a, b) show closely similar diagrams.
 \pretolerance=15000  \tolerance=15000 
 \lineskip=0pt \lineskiplimit=0pt
}
\end{figure}

\eject

      The pseudobulges that were postponed from Figure 5 are added to the BH--host correlations in Figure 7.  Hu (2008) was 
the first person to show that pseudobulges deviate from the $M_\bullet$\ts--\ts$\sigma_e$ correlation in having small BH masses.
This was confirmed with larger samples and extended to the $M_\bullet$\ts--\ts$M_{K,\rm bulge}$ and 
$M_\bullet$\ts--\ts$M_{\rm bulge}$ correlations by Greene \etal (2010) and by Kormendy, Bender, \& Cornell (2011).  
Figure 7 now shows this result for the largest available sample, that of KH13.

\vfill


 \includegraphics{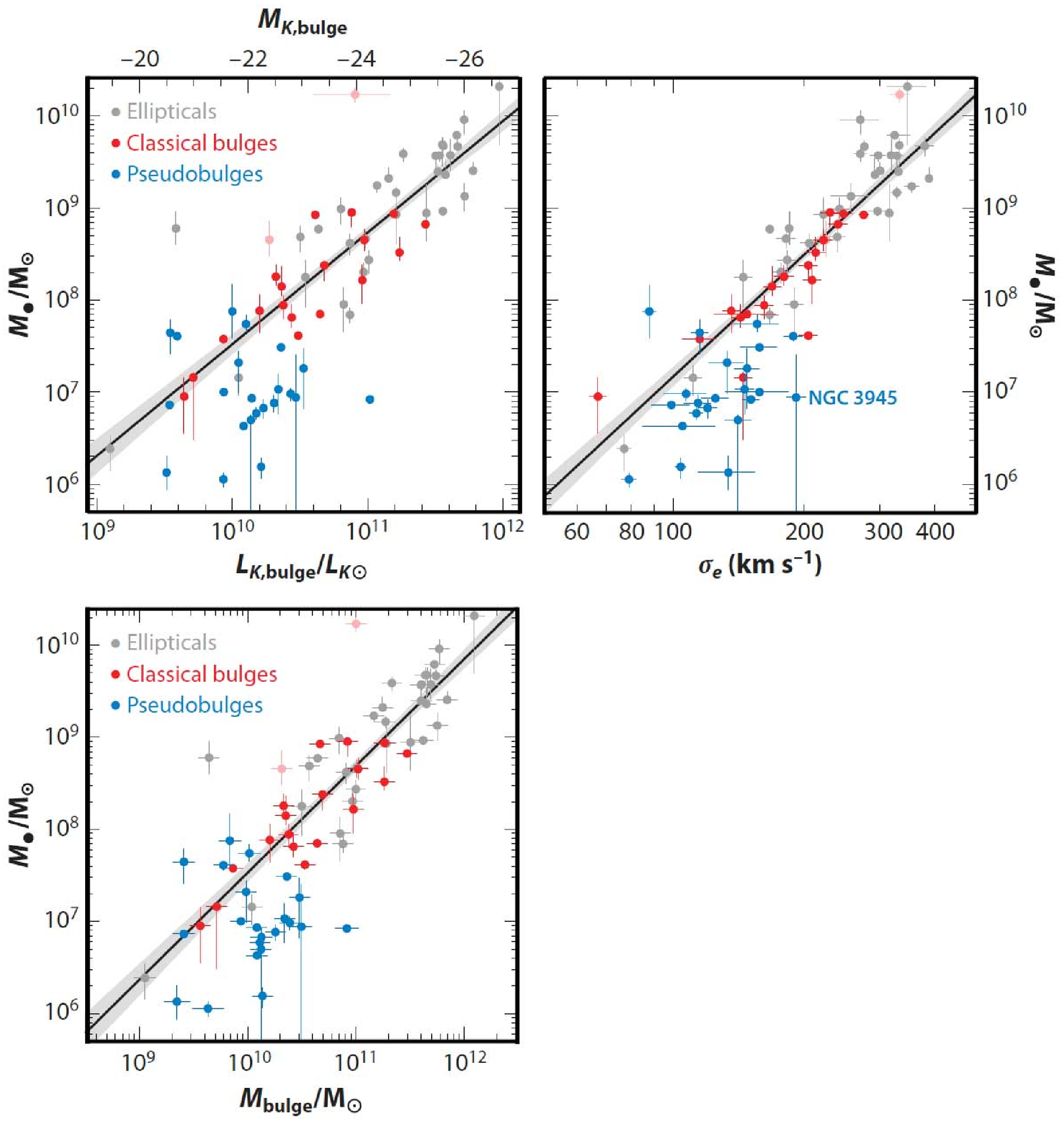}

\begin{figure}
\caption{
 \pretolerance=15000  \tolerance=15000 
 \lineskip=0pt \lineskiplimit=0pt
Correlations of BH mass with the $K$-band absolute magnitude and luminosity of the host bulge ({\it top-left panel}), with its
stellar mass ({\it bottom panel}), and with the mean velocity dispersion of the host bulge at radii that are large enough so that 
$\sigma_e$ is unaffected by the BH ({\it right panel}).  Gray points are for ellipticals, red points are for classical bulges, 
and blue points are for pseudobulges.  The lines with shaded 1-$\sigma$ uncertainties are symmetric least-squares fits to the 
classical bulges and ellipticals.  In all panels, pseudobulge BHs are offset toward smaller $M_\bullet$ from the correlations 
for classical bulges and ellipticals.  Absent any guidance from the red and gray points, we conclude the pseudobulge BHs do not 
correlate with their hosts in any way that is strong enough to imply BH-host coevolution.  From KH13, who 
tabulate the data and give sources.
}
\end{figure}

\eject

      Hints of this result are seen in McConnell \& Ma (2013); they compare early- and late-type galaxies and note
that many late-type galaxies have undermassive BHs.  This captures some of the result in Figure 7 but not all of it, 
because many S0 galaxies contain pseudobulges.  Similarly, Graham (2015) compares barred and unbarred galaxies and 
concludes that many barred galaxies have undermassive~BHs.  Again, this result is related to Figure 7 -- many (but 
not all) barred galaxies contain pseudobulges, and many (but not all) unbarred galaxies contain classical bulges.

      In Figure 7, the highest-$M_\bullet$ pseudobulge BHs largely agree with the correlations for classical
bulges and ellipticals; the lowest-$M_\bullet$ BHs deviate, but not by much more than an order of magnitude. 
Note that the BHs that we find in pseudobulges may be only the high-$M_\bullet$ envelope of a distribution that 
extends to much lower BH masses.  Still, why are pseudobulge BHs even close to the correlations? KH13  
argue that this natural: even one major merger converts a pseudobulge to a classical bulge, and then 
merger averaging manufactures an essentially linear correlation with a zeropoint near the upper end of the 
mass distribution of progenitors (see Figure\ts37 in KH13 and 
Peng 2007;
Gaskell 2010, 2011;
Hirschmann \etal 2010;
Jahnke \& Macci\`o 2011, who developed this idea).

      Turning next to disks: Figure 8 confirms the conclusion reached in Kormendy \& Gebhardt (2001) and
in Kormendy, Bender, \& Cornell (2011) that BH masses are completely uncorrelated with properties of their host disks.

\vfill



  \includegraphics{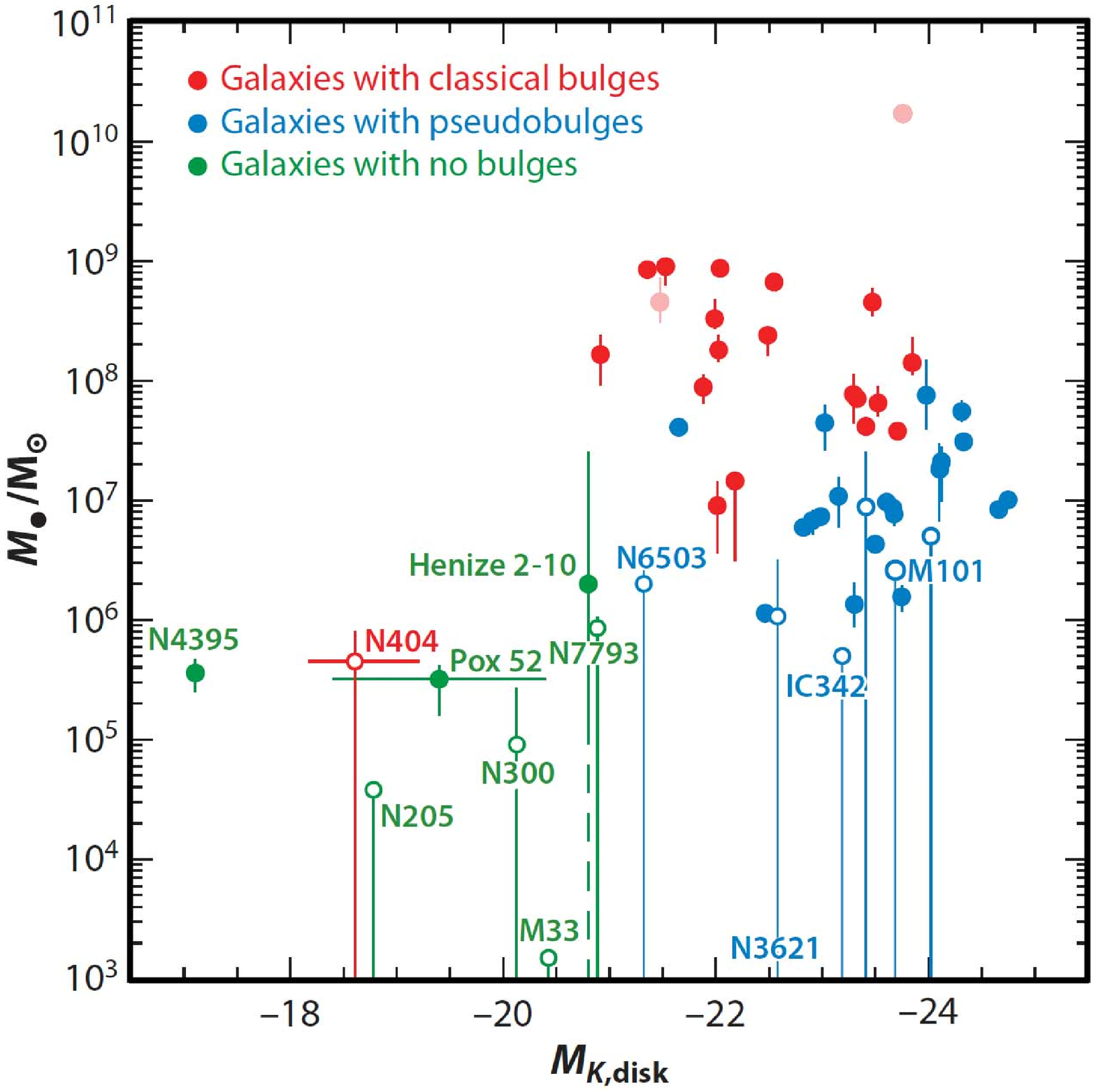}

\begin{figure}
\caption{
 \pretolerance=15000  \tolerance=15000 
 \lineskip=0pt \lineskiplimit=0pt
Black hole mass $M_\bullet$ vs $K$-band absolute magnitude of the disk of the host galaxy.  Filled circles are for galaxies 
with BH detections based on spatially resolved stellar or gas dynamics; open circles are for galaxies with upper limits on 
$M_\bullet$.  The strongest upper limit is $M_\bullet$ \lapprox 1500 $M_\odot$ in M{\ts}33 (Gebhardt \etal 2001).  Red and 
blue circles are for galaxies with classical and pseudo bulges, respectively.    Green points are for galaxies with no 
classical bulge and (almost) no pseudobulge but only a nuclear star cluster.  From KH13, who tabulate the
data and give sources.
}
\end{figure}

\eject

      M{\ts}33, with its strong upper limit on $M_\bullet$, briefly gave us the feeling that pure disks might not
contain BHs.  But it was clear all along that they can have AGNs.  Figure\ts8 includes bulgeless galaxies in which 
we find $10^{6 \pm 1}$-$M_\odot$ BHs.  The prototypical example is NGC 4395, a dwarf Sd galaxy with $M_V = -18.2$, with 
no classical or pseudo bulge, but with only a nuclear star cluster that has an absolute magnitude of 
$M_B \simeq -11.0$ and a velocity dispersion of \hbox{$\sigma$ \lapprox \ts30 $\pm$ 5 km s$^{-1}$} (Filippenko \& Ho 2003; 
Ho \etal 2009).  And yet, NGC 4395 is the nearest Seyfert 1 galaxy known (Filippenko \& Ho 2003).  It shows the  
signatures of BH accretion -- broad optical and UV emission lines (Filippenko, Ho \& Sargent 1993), variable X-ray 
emission (Shih, Iwasawa, \& Fabian 2003), and a compact, flat-spectrum radio core (Wrobel \& Ho 2006).  Peterson \etal (2005) 
get $M_\bullet = (3.6 \pm 1.1) \times 10^5 \, M_\odot$ by reverberation mapping.  This is the smallest BH mass  
measured by reverberation mapping.  But the BH in NGC 4395 is much more massive than $M_\bullet$ \lapprox \ts1500
$M_\odot$ in the brighter pure-disk galaxy M{\ts}33 ($M_V = -19.0$).

      This is the best example of many that are revealed in the observing programs of Ho, Barth, Greene, and collaborators 
and reviewed by Ho (2008) and by KH13.  Other important galaxies include Pox 52 
(Barth \etal 2004; 
Thornton \etal 2008)
and Henize 2-10 (Reines \etal 2011).  Broader AGN surveys to find low-mass BHs, many of them in late-type, pure-disk galaxies, include 
Greene \& Ho (2004, 2007),
Barth, Greene, \& Ho (2008), and
Dong \etal (2012).
The general conclusion is that classical and even pseudo bulges are not necessary equipment for the 
formation and nurture of supermassive BHs.

      We need one more result before we discuss implications for galaxy evolution:

      Very popular for more than a decade has been the suggestion that the fundamental correlation between BHs and
their host galaxies is not one with bulge properties but rather is a correlation with halo DM.  This was 
suggested by Ferrarese (2002) and supported by papers such as Baes \etal (2003).  The idea is attractive for galaxy 
formation theory, because then halo mass is the natural parameter to control AGN feedback (e.{\ts}g., Booth \& Schaye 2010).
The most robust part of our effort to model galaxy formation is the calculation of DM hiararchical clustering.  Conveniently,
DM mass is then provided by halo-finder algorithms.  

      However, we can now be confident that {\it halo DM does not correlate directly with $M_\bullet$ independent
of whether or not the galaxy contains a bulge} (Kormendy \& Bender 2011).  This result is reviewed in detail and 
with the largest  galaxy sample in KH13. They list eight arguments against Ferrarese's conclusion.
Some are based on examining the proxy parameters that she used to make her arguments ($\sigma$ for $M_\bullet$
and $V_{\rm circ}$ for the DM; e.{\ts}g. we now know that $\sigma$ is not a proxy for BH mass for pseudobulge galaxies:
Figure 7 here).  Some arguments are based on the direct correlation of measured $M_\bullet$ with $V_{\rm circ}$:
there is essentially no correlation unless the galaxy has a classical bulge.  Perhaps the most telling argument is based
on the well determined relationship between the stellar mass $M_*$ and the DM mass $M_{\rm DM}$ of galaxies.  
Behroozi, Wechsler, \& Conroy~(2013) show that $M_*/M_{\rm DM}$ reaches a maximum at $M_{\rm DM} \simeq 10^{12}$ $M_\odot$
and is smaller at both higher and lower $M_{\rm DM}$ (see also Fig.~9 here).  Together with the correlation (Equation 5) between $M_\bullet$
and $M_{\rm bulge} \simeq M_*$ (exact for ellipticals and approximate for bulge-dominated galaxies), Behroozi's
result implies that the relationship between $M_\bullet$ and $M_{\rm DM}$ is complicated,
\vskip -7pt
$$ M_\bullet \propto M_{\rm DM}^{2.7}~~~{\rm at}~~M_{\rm DM}  \ll 10^{12}{\ts}M_\odot~~,   \eqno{(6)}$$ 
\vskip -9pt
\noindent but
\vskip -9pt
$$ M_\bullet \propto M_{\rm DM}^{0.34}~~~{\rm at}~~M_{\rm DM} \gg 10^{12}{\ts}M_\odot~, \eqno{(7)}$$
\vskip -1pt
\noindent with a kink in the correlation at $M_{\rm DM} \simeq 10^{12}$\ts$M_\odot$. 
Meanwhile, the $M_\bullet$\ts--\ts$M_{\rm bulge}$ correlation is log linear with small scatter from the lowest to the 
highest bulge masses in Figure 5.  This correlation shows no kink at $M_{\rm DM} \sim 10^{12}$ $M_\odot$ corresponding 
to $M_{\rm bulge} \sim 3 \times 10^{10}$ $M_\odot$ (see Figure 7).  The simplicity of $M_\bullet$\ts--\ts$M_{\rm bulge}$
versus the complexity of $M_\bullet$\ts--\ts$M_{\rm DM}$ is another argument in favor of the
conclusion that BHs coevolve with bulges and ellipticals but not directly with DM halos.


\vskip -26pt

\centerline{\phantom{00000000000000000}}

\subsection{AGN Feedback and the Coevolution (Or Not) of Supermassive Black Holes and Host Galaxies}

      Implications for the coevolution (or not) of BHs and host galaxies are reviewed by Kormendy \& Ho (2013).  
They distinguish four modes of AGN feedback:

\begin{enumerate}

\item[(1){\kern -3pt}]{Galaxies that are not dominated by classical bulges -- even ones like NGC 4736 that contain big
                       pseudobulges -- can contain BHs, but these grow by low-level AGN activity that involves too little energy 
                       to affect the host galaxy.  Whether or not AGNs are turned on when we observe them, these galaxies 
                       actively form stars and engage in secular evolution by the redistribution of gas.
                       Most AGNs at $z \sim 0$ and probably out to $z \sim 2$ are of this kind.  They include giant galaxies
                       such as our Milky Way, with outer circular-orbit rotation velocites $V_{\rm circ} > 220$ km s$^{-1}$.
                       These galaxies are not correctly described by simple prescriptions in which gravitational potential 
                       well depth controls AGN feedback.}

\item[(2){\kern -3pt}]{Most consistent with the prevailing emphasis on AGN feedback are classical bulges and
                       coreless-disky-rotating ellipticals.  They satisfy the tight correlations between $M_\bullet$
                       and bulge properties in Figure 5.  It is likely (although the engineering is not fully understood) that AGN
                       feedback helps to establish these $M_\bullet$--host relations during dissipative (``wet'') major mergers.
                       This must happen mostly at high $z$, because gas fractions in major mergers at $z \sim 0$ are small,
                       and indeed, mergers in progress at $z \simeq 0$ do not satisfy the $M_\bullet$ correlations.
                       It is important to note that even small Es with $V_{\rm circ}$ \lapprox \ts100 km s$^{-1}$ (e.{\ts}g., 
                       M{\ts}32) satisfy the $M_\bullet$--host correlations, whereas even giant pure disks (e.{\ts}g., M{\ts}101)
                       do not.  Coevolution is not about potential well depth.  Coevolution (or not) is determined by whether
                       (or not) the galaxy contains a classical bulge of elliptical -- i.{\ts}e., the remnant of at least one
                       major merger.
                       }

\item[(3){\kern -3pt}]{The highest-mass ellipticals are coreless-boxy-nonrotating galaxies whose most recent 
                       mergers were dissipationless (``dry'').  These giant ellipticals inherit any feedback magic --
                       including the $M_\bullet$--host relations -- from (2).  In them, AGN feedback plays a different,
                       essentially negative role.  It keeps galaxy formation from ``going to completion'' by keeping 
                       baryons suspended in hot gas.  With masses $M > M_{\rm crit}$ in Section 7, these galaxies hold 
                       onto hot, X-ray-emitting gas that is believed to prevent cold-gas dissipation and to quench star
                       formation.  However, X-ray gas cooling times are short, 
                       and so -- given that we observe only weak temperature gradients -- something
                       must keep the hot gas hot.  One such process is gas infall from the cosmological web (Dekel \&
                       Birnboim 2006).  Another is ``maintenance-mode AGN feedback'' (see Fabian 2012 for a review).  
                       All proposed heating processes may be important.  See Section\ts7.
                       \pretolerance=100000\tolerance=100000}

\item[(4){\kern -3pt}]{The averaging that is inherent in galaxy mergers may significantly decrease the scatter in the
                       $M_\bullet$--host correlations.  That is, during a merger, the progenitors' stellar masses add and so
                       do their BH masses.  In the absence of new star formation, the effect is to decrease the correlation
                       scatter.  Recall a conclusion in Section 4 that only a modest amount of star formation happens during
                       mergers.  So the central limit theorem ensures that the scatter in BH correlations 
                       with their hosts decreases as $M_\bullet$ increases via either wet or dry mergers.}

\end{enumerate}

      In summary, KH13 provides the largest available database on BH detections via spatially
resolved dynamics, putting the many heterogeneous discovery papers on a homogeneous system of (for example)
distances and magnitudes, and incorporating many $M_\bullet$ corrections from the recent literature.  Homogeneous
data are also provided for all BH host galaxies, including all disk-galaxy hosts, many of which had not
previously been studied.  Bulge-pseudobulge classifications are provided based on multiple classification
criteria (cf.{\ts}Section\ts2.1 here), and (pseudo)bulge-disk photometric decompositions are derived for all galaxies
that did not previously have photometry.  The results (their Tables~2~and~3) are an accurate enough database
to allow Kormendy \& Ho (2013) to derive a number of new conclusions about BH-host correlations and their
implications.  Some of these are reviewed above.  Others, such as correlations (or not) with nuclear star clusters
and globular cluster systems, are omitted here, in part to keep the length of this paper manageable, and in part
because the connection with galaxy bulges is less direct than it is for subjects that we cover. 

      Many of our conclusions disagree with Graham (2015).  Within the subjects that I have reviewed in this paper, 
I have tried to explain why.  Readers are encouraged to compare the accuracy of our data sets 
(particularly $M_\bullet$ measurements), our results, and the physical picture in which they are embedded.  
We believe that the observational conclusions reached in KH13 are robust, and the essential implications 
for galaxy evolution -- the big picture of what happens, if not the engineering details -- are well established.  
Section 7 is an important example.

\vskip -25pt

\centerline{\phantom{00000000000000000}}

\section{Quenching of Star Formation}

      Many papers on star formation histories begin by setting up a ``straw-man target'' that the quenching of star formation is mysterious.  
In contrast, it strikes me that the literature shows encouraging convergence on a picture at least at $z < 1$ in which~well defined 
processes convert ``blue cloud'' star-forming galaxies to ``red sequence'' red and dead galaxies.  This section rephrases Section 6.2 
to describe this picture.

      The essential observation that has driven progress on this subject is summarized in Figure 9.  The left panel shows the
Allen \etal (2011) version of the Behroozi~\etal (2013) result that led to Equations (6) and (7) in Section 6.1.  I use
it because the abscissa is in the same units as in the right panel.  It shows that the ratio of stellar mass to total
mass reaches a maximum at $V_{\rm circ} \sim 300${\ts}km{\ts}s$^{-1}$ or, in Behroozi \etal (2013), at $M_{\rm DM} \sim 10^{12}$ $M_\odot$.
This maximum is $\sim$\ts1/5 of the cosmological baryon fraction, so most baryons in the universe have not yet made stars.
\hbox{Lower-mass} halos have smaller stellar fractions ({\it left panel\/}) and smaller baryon fractions ({\it right panel\/}) because -- we believe --
the baryons have increasingly been ejected from DM halos by star-formation and supernova feedback or never accreted after cosmological
reionization.  But the focus here is on higher DM masses.  They, too, have smaller stellar mass fractions than at the ``sweet spot''
halo mass of $10^{12}$ $M_\odot$.  But Figure\ts9 ({\it right\/}) shows that these baryons are not ``missing'' at $M_{\rm DM} \gg 10^{12}$\ts$M_\odot$.
On the contrary, the total baryon fraction converges to essentially the cosmological value in the highest-mass halos, which are
halos of rich clusters of galaxies.  This is the by-now well known result that, as $M_{\rm DM}$ grows above $10^{12}$ $M_\odot$ and $V_{\rm circ}$
grows above 300 km s$^{-1}$, an increasingly large fraction of the baryons are indeed present but have not made stars.  Rather, they are suspended 
in hot, X-ray-emitting gas, until in rich clusters of galaxies, that hot gas outmasses the stellar galaxies in the cluster by $1.0 \pm 0.3$ dex 
(Kravtsov \& Borgani 2012).  This has led to the essential idea of \hbox{``$M_{\rm crit}$ quenching''} of star formation by X-ray-emitting gas, 
which can happen provided that the DM mass is larger than the critical mass, $M_{\rm DM}$ \gapprox \ts$M_{\rm crit} \simeq 10^{12}$\ts$M_\odot$,
that is required to support the formation and retention of hot gas halos (e.{\ts}g.,
Birnboim \& Dekel 2003;
Kere\v s \etal 2005;
Cattaneo \etal 2006, 2008, 2009;
Dekel \& Birnboim 2006, 2008;
Faber \etal 2007;
KFCB;
Peng \etal 2010, 2012;
KH13,
Knobel \etal 2015, and
Gabor \& Dav\'e 2015).

\vfill


 \includegraphics{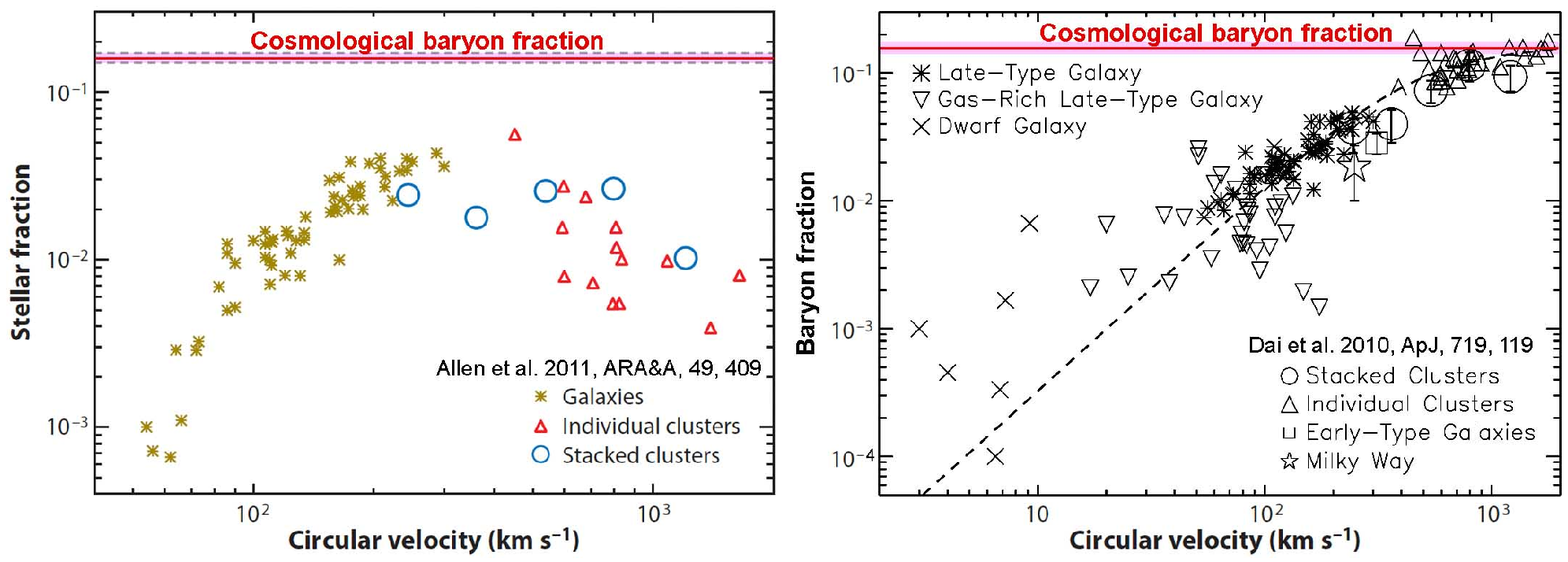}

\begin{figure}
\caption{
 \pretolerance=15000  \tolerance=15000 
 \lineskip=0pt \lineskiplimit=0pt
Stellar mass fraction $M_*/(M_{\rm baryon} + M_{\rm DM})$ ({\it left\/}) and total baryon mass fraction
$M_{\rm baryon\/}/(M_{\rm baryon} + M_{\rm DM})$ ({\it right\/}) versus a circular-orbit rotation velocity
$V_{\rm circ} \sim \sqrt{G M_{\rm DM} / r}$ (Dai \etal 2010) that approximately characterizes the total
mass distribution.  Here $M_*$ is the stellar mass, $M_{\rm DM}$ is the DM halo mass, $r$ is the radius of
the halo, and $G$ is the gravitational constant.  The cosmological baryon fraction has been adjusted very
slightly to $0.16 \pm 0.01$, i.{\ts}e., the mean of the WMAP and Planck measurements (Hinshaw \etal 2013 and
Planck Collaboration 2014, respectively).  Both figures originally come from Dai \etal (2010).
}
\end{figure}

\eject

      The transition mass between galaxies that should contain X-ray gas and those that should not is consistently derived 
by a variety of theoretical arguments and is consistent checked via a variety of observational tests.  It should occur at the 
DM mass at which the hot gas cooling time is comparable to the infall time (Rees \& Ostriker 1977).  Birnboim \& Dekel (2003) and Dekel \& 
Birnboim (2006, 2008) argue from theory and Kere\v s \etal (2005) find from SPH simulations that gas that is accreted during hierarchical 
clustering falls gently into shallow potential wells and makes star-forming disks, whereas gas crashes violently onto giant galaxies
and is shock-heated to the virial~temperature.  It is this hot gas that quenches star formation.  Calculated hot-gas cooling times are 
short; this led to the well known ``cooling flow problem'' (Fabian 1994).~\hbox{But X-ray} measurements of temperature profiles 
now show that they are much shallower than cooling-time calculations predict in the absence of heating 
(McNamara \& Nulsen 2007;
Kravtsov \& Borgani 2012;
Fabian 2012).  
Debate continues about how the gas is kept hot; 
Dekel \& Birnboim (2006, 2008) 
suggest that the required heating is caused by continued accretion; AGN feedback is another candidate (e.{\ts}g.,
Best et al. 2006; 
Best 2006, 2007a, b;
Fabian 2012;
Heckman \& Best 2014),
and dying stars return gas to the intergalactic medium at just the right kinetic temperature (Ostriker 2006).
The engineering details need to be sorted out.  It is likely that all processes are important.   But from 
the point of view of this paper, the engineering is secondary.  The important point is that the galaxies and clusters 
tell us that they know how to keep the gas hot.

      Many observed properties of galaxies can be understood in the context of $M_{\rm crit}$ quenching.  E.{\ts}g.,
it allows semianalytic models of galaxy formation to reproduce the color bimodality of galaxies (``red sequence'' versus ``blue cloud'';
Blanton \& Moustakas 2009) as a function of redshift (Cattaneo et al. 2006, 2008, 2009). 

      Faber \etal (2007) and KFCB emphasize the connection of the above results to this paper: {\it $M_{\rm crit}$ star-formation 
quenching is believed to explain the difference between the two kinds of ellipticals discussed in Section 4.1.1.}   I noted there that 
classical bulges and coreless-disky-rotating ellipticals generally do not contain X-ray-emitting gas, whereas core-boxy-nonrotating 
ellipticals contain more X-ray gas as their luminosities increase more above $L_{\rm crit} = 10^{10.2}$ $L_{B\odot}$~(Figure\ts3).  Now, 
$L_{\rm crit}$ corresponds to $M_V \simeq -20.9$; i.{\ts}e., 0.6 mag fainter~than~the~divide between coreless-disky-rotating
and core-boxy-nonrotating ellipticals.  This is a factor of almost 2.  If the most recent event that made an elliptical
was an \hbox{equal-mass} merger, then {\it the divide betweeen coreless-disky-rotating and core-boxy-nonrotating~ellipticals happens at a 
luminosity below which neither of the merger progenitor galaxies should have contained \hbox{X-ray} gas and above which one or both progenitor 
galaxies should have contained \hbox{X-ray} gas.  Thus KFCB point out~that~the \hbox{E{\ts}--{\ts}E} dichotomy occurs at the correct luminosity 
so that coreless-disky-rotating ellipticals formed in wet mergers whereas core-boxy-nonrotating ellipticals formed in dry mergers.}

      Specifically,  $M_V \simeq -20.7$ for merger progenitors corresponds (using $M/L_V \sim 6$) to a stellar mass of $M_* \simeq 1 \times 10^{11}$
$M_\odot$ or, using a baryon-to-total mass ratio of 1/6 (Komatsu \etal 2009), to $M_{\rm DM} \simeq 6 \times 10^{11}$ $M_\odot$.
And the divide between coreless-disky-rotating Es and core-boxy-nonrotating Es happens at $M_{\rm DM} \simeq 10^{12}$ $M_\odot$.
So the agreement with the above picture of $M_{\rm crit}$ star-formation quenching is good.  

      Thus our picture of the formation of classical bulges and elliptical galaxies by wet and (at $M_{\rm DM} > 10^{12}$\ts$M_\odot$) dry
major mergers (Section 4 of this paper) is a tidy addition to our developing paradigm of star-formation quenching.  Many details of the 
structure of classical bulges and ellipticals (e.{\ts}g., the list in Section 4.1.1) fit into and support this paradigm.  
But the paradigm is more general than just an explanation of the E{\ts}--{\ts}E dichotomy.  I turn to these more general aspects next:

      In a seminal paper, Peng \etal (2010) use a few robust observations~to~derive very general conclusions about how quenching must work.
They do this completely operationally, without any need to identify the physical mechanism(s) of quenching.  At redshift $z$\ts$\sim$\ts0 
(Sloan Digital Sky Survey) and out to $z$\ts$\sim$\ts1 (zCOSMOS survey: Lilly \etal 2007) the most essential observations used are (1) that the specific star 
formation rate is almost independent of galaxy mass (there is a ``main sequence'' of star formation) but with rapidly decaying specific star 
formation rate as $z \rightarrow 0$, and (2) that star-forming galaxies satisfy a Schechter (1976) mass function whose characteristic mass is 
almost independent of $z$.  From a discussion of how star formation operates to reproduce the above and other observations, 
they deduce that quenching is driven by galaxy mass and by galaxy environment and that these two modes (not identified physically) are separable
and independent.  Plus there must be an additional quenching mode that is associated with bulge formation via mergers.  
Figure 10 connects their picture with the quenching paradigm that we review here.

\vfill


 \includegraphics{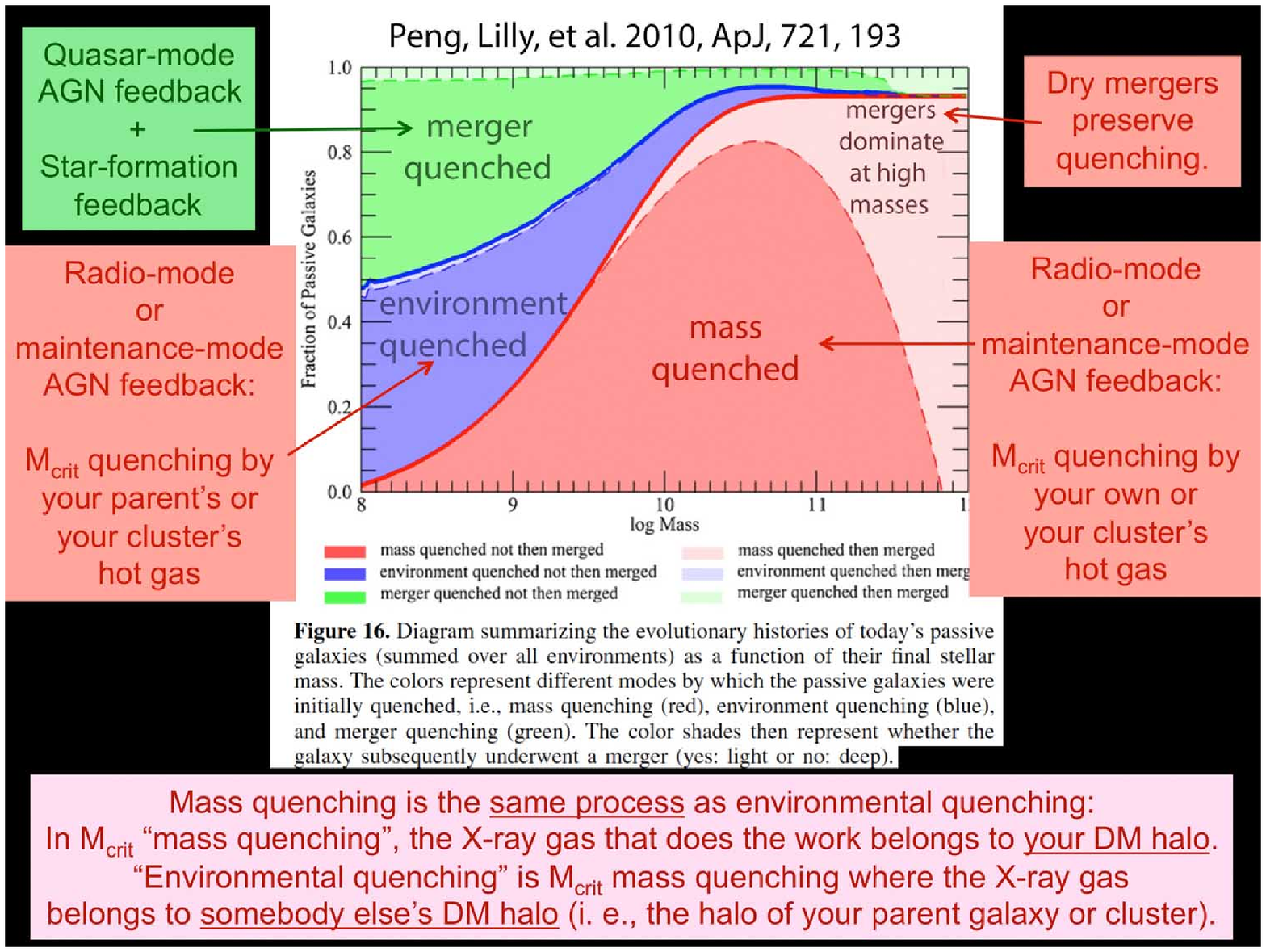}

\begin{figure}
\caption{
 \pretolerance=15000  \tolerance=15000 
 \lineskip=0pt \lineskiplimit=0pt
Powerpoint slide connecting the star-formation quenching picture of Peng \etal (2010: central figure and its caption)
with the picture that is summarized in this paper (surrounding text).   
}
\end{figure}

\eject

      Peng \etal (2010) emphasize that their analysis is operational:~it~identifies the conditions in which quenching 
must operate, but it does not identify quenching mechanisms.  However, with this section's background 
on $M_{\rm crit}$ quenching and with results from KH13 on BH{\ts}--{\ts}host-galaxy coevolution (or not), 
we can identify aspects of our developing physical picture of star-formation quenching with the conclusions of Peng \etal (2010). 
This is illustrated in Figure 10.

      The masses used in Peng \etal (2010) are estimated by integrating star formation rates and by fitting spectral energy distributions;
in essence, they are stellar masses.  Figure 10 suggests that mass quenching tends to happen at masses $\sim 10^{10.5}$\ts$M_\odot$.  
In Figure 7 of Peng \etal (2010), the fraction of quenched galaxies (independent of environment) reaches 50\ts\% at $\sim 10^{10.6}$\ts$M_\odot$ 
and 80\ts\% at $\sim 10^{11.25}$\ts$M_\odot$.  These correspond to $M_{\rm DM} \sim 10^{11.4}$ to $10^{12}$ $M_\odot$.  The good agreement with 
$M_{\rm crit}$ suggests that Peng's ``mass quenching'' is precisely our ``$M_{\rm crit}$ quenching'' by hot gas.

      Peng \etal (2010) conclude further that some low-mass galaxies are quenched by their environments.  That is, these galaxies are
quenched because they are satellites of higher-mass objects -- ones (either individual galaxies or clusters of galaxies) that 
{\it can have\/} masses $M_{\rm DM}$ \gapprox $M_{\rm crit}$.  {\it I suggest that Peng's ``environmental quenching''
is the same physical process as mass quenching, but in Peng's mass quenching, the X-ray gas that does the work belongs to the galaxy that
is being quenched, whereas in environmental quenching, the X-ray gas that does the work belongs to somebody else; i.{\ts}e., to
the quenched galaxy's parent giant galaxy or galaxy cluster.}  This idea is verified by Peng \etal (2012), Knobel \etal (2015), and
Gabor \& Dav\'e (2015).

      The suggested connection with KH13 then is this: Both mass and environment quenching are aspects of
point 3 in Section 6.2 -- they are effects of hot gas that is kept hot by a combination of maintenance-mode AGN feedback and other processes 
such as continued infall of gas from the cosmological hierarchy and the injection of the kinetic energy of gas that is shed by dying stars.

      But the above quenching processes are not sufficient.  It is easy~to~explain~why\ts-- to give an example that mass quenching and
environment quenching cannot explain.  {\it What quenches field S0 galaxies with masses $M$ $\ll$\ts$M_{\rm crit}$?}  Kormendy \& Ho (2013) 
suggest that they are quenched in the context of wet galaxy mergers that include starbursts, with energy feedback from the starburst beginning 
the job of quenching and AGN feedback (Section 6.2, point 2) finishing the job.  It seems natural to suggest that this is the Peng's
``merger quenching''.  Observations of gas outflows~in high-$z$, star-forming galaxies such as submillimeter galaxies{\ts}--{\ts}at least some
of which are mergers{\ts}--{\ts}are reviewed in KH13.  Of course, bulge-formation and $M_{\rm crit}$ quenching can be mutually
supportive (e.{\ts}g., Woo \etal 2015).

      Once star formation is quenched at $M > M_{\rm crit}$, then dry mergers preserve both the quenched state and the $M_\bullet${\ts}--{\ts}host
correlations (Section 6.2, point 4 and modes ``mass quenched then merged'', ``environment quenched then merged'', and ``merger quenched
then merged'' in Figure 10).

      The biggest remaining question in our $z < 1$ picture is this: {\it In merger-quenched galaxies that have $M \ll M_{\rm crit}$, i.{\ts}e.,
in objects in which X-ray gas is not available even after the merger is finished, what preserves the quenched, red and dead state?
We do not know, but episodic, low-level AGN feedback may be~the~answer.}

\vfill

      The biggest overall uncertainty is that quenching may operate \hbox{differently at $z$ \gapprox {\kern 0.6pt}2}.  Dekel and Birnboim argue 
(1) that $M_{\rm crit}$ is higher at high $z$, when gas fractions in galaxies and gas accretion rates onto galaxies~are~both~higher~and 
(2) that cold streams can penetrate hot gas at high $z$ and contribute to the growth of disks at masses that are unattainable at $z \sim 0$
(Dekel \etal 2009).
Another difference involves the observation that most star-forming galaxies define a main sequence of star formation with few outliers,
implying that duty cycles are long and hence that star formation is not driven primarily by short-duration events such as mergers (Section 4.1.5).  
When strong gas outflows are seen in star-forming galaxies at $z \sim 2$, the inference is that some combination of star formation and AGN
feedback is responsible but that these are not primarily driven by major mergers (e.{\ts}g.,
F\"orster Schreiber \etal 2014; 
Genzel \etal 2014).
Because these processes are also associated with bulge growth in disk galaxies (Lang \etal 2014), the most consistent interpretation
that also includes the $M_\bullet$ correlation results is that the bulge growth in these objects is by clump cluster sinking (Section 3 here).
Genzel (private communication) suggests that Peng's mass quenching may be this outflow process associated with more-or-less steady-state star formation,
AGN feedback, and classical bulge growth.  On the ``plus side'', there is clearly a danger that our tidy $z < 1$ picture is basically 
correct but not a description of what happens at $z \gg 1$.  On the other hand, we already know that many details of galaxy structure are well
explained by the $z \sim 0$ picture.  Particularly important is the natural explanation of cores in dry-merger remnants and central extra light
in wet-merger remnants (see KFCB).  Alternative suggestions for quenching mechanisms at high $z$ have not addressed and solved the 
problem of also explaining these aspects of $z \sim 0$ galaxy structure.  This is not a proof that the suggested high-$z$ processes
are wrong.

      It seems reasonable to conclude that our $z < 1$ picture of star formation quenching is robust.  Mostly, it needs clarification of 
engineering details.  In marked contrast, star formation quenching at $z$ \gapprox 2 is less well understood, although progress is rapid.

\section{A Partial Summary of Outstanding Problems}

      I conclude with a summary of the most important outstanding problems.  I restrict myself
to big-picture issues and do not address the myriad engineering details that are unsolved
by our present state of the art.  They are, of course, vitally important.  But a comprehensive list 
would require a paper of its own.  I therefore refer readers to earlier chapters of this book, 
which discuss many of these problems in detail.

\begin{enumerate}

\item[(1){\kern -3pt}]{I emphasized in Section 4.1.3 that, to me, the most important goal is to produce
                       realistic classical bulges\ts$+${\ts}ellipticals and realistic disks that overlap over
                       a factor of $>$ 1000 in mass but that differ from each other in
                       ways that we observe over the whole of this range.  They can combine  
                       with any $B/T$ from 0 to 1, but the differences between bulges
                       and disks depend very little on $B/T$.
                      }

\item[(2){\kern -3pt}]{Four decades of work on $z \simeq 0$ galaxies showed convincingly that major mergers
                       convert disks into classical bulges and ellipticals with the observed~properties,
                       including S\'ersic index, fundamental plane parameter correlations, 
                       intrinsic shape and velocity distributions, both as functions of mass,
                       the presence of cores or central extra light, and isophote shape.  This work also
                       suggested that merger rates were higher in the past, and modern observations confirm
                       this prediction.  By the mid-1990s, we had converged on a picture in which classical
                       bulges and ellipticals were made in major mergers.  Enthusiasm for mergers was probably
                       overdone, but now, the community is overreacting in the opposite direction.  The 
                       successes of the 1970s--1990s are being forgotten, and -- I believe -- we 
                       have come to believe too strongly that minor mergers control galaxy evolution.  Reality
                       probably lies between these extremes.  For today's audience, the
                       important comment is this: The observations that led to our picture of E formation via 
                       major mergers have not been invalidated.  I suggest that the profitable way forward is
                       to use what we learn from $z \simeq 0$ mergers-in-progress to explore how mergers
                       make bulges and ellipticals at higher $z$, including (of course) differences 
                       caused (for example) by large gas fractions and including new ideas, such as violent
                       disk instabilities that make clumps that make bulges.  For this still-elusive
                       true picture, it is OK that mergers are rare, because ellipticals are rare, too, and classical
                       bulges are rarer than we thought.  And it is OK that most star formation does not 
                       happen in mergers, because ellipticals are rare anyway, and because their main bodies
                       are made up of the scrambled-up remnants of already-stellar progenitor disks.
                      }

\item[(3){\kern -3pt}]{The most important unsolved problem is this: How did hierarchical clustering produce so many giant
                       galaxies (say, those with $V_{\rm circ}$ \gapprox 150 km s$^{-1}$) with no sign of a 
                       classical bulge?  This problem is a very strong function of environment -- in field
                       environments such as the Local Group, most giant galaxies are bulgeless, whereas in
                       the Virgo cluster, most stars live in classical bulges and elliptical
                       galaxies.  The clue therefore is that the solution involves differences in accretion
                       (gentle versus violent) and not largely internal physics such as star-formation or AGN feedback.
                      }

\item[(4){\kern -3pt}]{Calculating galaxy evolution {\it ab inito\/}, starting with $\Lambda$CDM density fluctuations,
                       constructing giant $n$-body simulations of halo hierarchical clustering, and then adding baryonic
                       physics is the industry standard today and the way of the future.  It is immensely difficult
                       and immensely rewarding.  It is not my specialty, and I have only one point to add to the
                       excellent review by Brooks \& Christensen: {\it Observations hint very strongly that we put
                       too much reliance on feedback to solve our engineering problems in producing realistic galaxies.
                       Observations of supermassive BH demographics tell us that AGN feedback
                       does not much affect galaxy structure or star formation until mergers start to make classical
                       bulges.  And point (3) emphasizes that environment and not gravitational potential well depth
                       is the key to solving the problem of giant, pure-disk galaxies.} 
                      }

\item[(5){\kern -3pt}]{We need to fully integrate our picture of disk secular evolution into our paradigm of galaxy evolution.
                       As observed at $z \simeq 0$, this picture is now quite detailed and successful.  Essentially all of 
                       the commonly occurring morphological features of galaxies -- bars, (nuclear, inner, and outer) rings,
                       nuclear bars, and pseudobulges -- are at least qualitatively explained within this picture.  Some of
                       these details are beyond the ``targets'' of present galaxy-formation simulations.  But pseudobulges
                       are immediately relevant, because our recognition of them has transformed our opinions about
                       classical bulges.  They are much rarer than we thought.  In particular, small classsical bulges
                       are {\it very\/} rare.  And although some galaxies have structure that is completely determined
                       by the physics of hierarchical clustering, others -- and they dominate in the field -- appear to have
                       been structured almost exclusively by secular processes.  Incorporating these processes is a challenge, because 
                       slow processes are much more difficult to calculate than rapid processes.  But secular evolution is an
                       ideas whose time has come (Sellwood 2014), and we need to include it in our paradigm.
                      }

\item[(6){\kern -3pt}]{At the same time, our {\it quantitative\/} understanding of secular evolution needs more work.  For
                       example, we need a study similar to Dressler's (1980) work on the morphology-density relation:
                       We need to measure the luminosity and mass functions of disks, pseudobulges, and classical bulges$+$ellipticals,
                       all as functions of environmental density.  At present, we have essentially
                       only two ``data points'' -- the extreme field (Kormendy \etal 2010; Fisher \& Drory 2011) and
                       the Virgo cluster (see Kormendy \etal 2010).  This is already enough to lead to point (3) in
                       this list.  We need corresponding studies in more environments
                       that span the density range from the field to the richest clusters.  This will not be easy,
                       first because we need high spatial resolution whereas observing more environments drives us to 
                       larger distances, and second because of point (7).
                      }

\item[(7){\kern -3pt}]{Our picture of disk secular evolution predicts that many galaxies should contain
                       both a classical and a pseudo bulge.  Work on the subject has concentrated on 
                       extremes -- on galaxies that are dominated by one kind of bulge or the
                       other.  Samples of large numbers of galaxies will inevitably have to face 
                       the challenge of separating at least three components (bulge, pseudobulge, and disk) and 
                       in many cases more (bar, lens, \dots).  We also need to be able to find pseudobulges
                       in face-on barred galaxies (see Section 2.1).  But it is easy to overinterpret details
                       in the photometry.  The best way to approach this problem is probably to begin with infrared
                       observations of nearly-edge-on galaxies (e.{\ts}g., Salo \etal 2015).
                      }

\item[(8){\kern -3pt}]{Are classical bulges really indistinguishable from ellipticals?  The structural
                       parameter scaling relations shown in Figure 4 (based on many authors' work) show that
                       they are closely similar. I use this result thoughout the present~paper.  It is central
                       to Renzini's (1999) paraphrase of the classical morphological definition: ``A bulge is nothing 
                       more nor less than an elliptical galaxy that happens to live in the middle of a disk.''
                       But not everybody agrees.  Based on multi-component decompositions, 
                       different fundamental plane correlations for classical bulges and ellipticals have been found by
                       Gadotti (2008,~2009,~2012) and by
                       Laurikainen \etal (2010).
                       We need to resolve these differences.~At~stake is an understanding of whether classical  bulges
                       and ellipticals form\ts--{\ts}as~I~suggest\ts-- by essentially the same major merger process
                       or whether important variations in that process produce recognizably different results.  In particular,
                       it is not impossible that we can learn to distinguish ellipticals and perhaps {\it some\/} bulges
                       that form via mergers of distinct galaxies from other bulges that form via the mergers of 
                       mass clumps that form in unstable disks.  Both processes drive additional gas toward the center,
                       but it is possible that bulge formation via disk instabilities is intrinsically
                       more drawn out in time with the result (for example) that ``extra light components'' such as
                       those studied in Kormendy (1999), KFCB, and Hopkins \etal (2009a) are smoothed away and 
                       unrecognizable in the resulting classical bulges but not in disky-coreless-rotating ellipticals.
                      } 

\item[(9){\kern -3pt}]{Returning to elliptical galaxies: KFCB present a detailed observational picture and ARA\&A-style
                       review of the two kinds of ellipticals in large part as seen in the Virgo cluster.  Hopkins \etal
                       (2009a, b) present modeling analyses of wet and dry mergers, respectively. {\it We need to know how
                       this very clean picture as seen in the nearest rich cluster translates into other environments.}
                       Much of the work published by Lauer \etal (1995, 2005, 2007a, b), by Faber \etal (1997), by
                       Kormendy \& Bender (1996, 2013), and by Bender \etal (1989) applies to broader ranges of 
                       environments.  It suggests that the picture summarized here in Section 4.1.1 is basically
                       valid but that the distinction between coreless-disky-rotating and core-boxy-nonrotating galaxies 
                       is somewhat ``blurred'' in a broader range of environments.  For example, $M_V = -21.6$
                       cleanly separates the two kinds in Virgo, with only one partial exception (NGC 4621 at 
                       $M_V = -21.54$ has $n = 5.36^{+0.30}_{-0.28}$ characteristic of core galaxies, but it has
                       a small amount of extra light near the center).  However, the above papers and others show
                       that the two galaxy types overlap over a range of absolute magnitudes from about $M_V = -20.5$
                       to about $M_V = -23$.  In the overlap range and occasionally outside it, some classification
                       criteria in Section 4.1.1 conflict with the majority.  We should not be surprised that heterogeneous
                       formation histories can have variable outcomes; on the contrary, it is encouraging to see
                       as much uniformity as we see.  Still, a study of how the systematics depend on environment
                       should be profitable.
                      } 

\item[(10){\kern -3pt}]{Still on ellipticals and classical bulges: The SAURON and ATLAS$^{\rm 3D}$ teams have
                       carried out an enormous amount of truly excellent work on nearly all aspects of bulge$+$E
                       structure and evolution.  A review is in preparation by Cappellari~(2015).
                       {\it It is natural to ask how the picture of bulges and ellipticals developed by the SAURON
                       and ATLAS$^{\rm 3D}$ papers compares with the one outlined in Section 4 here.  The answer is that
                       they agree exceedingly well.}  There are differences in emphasis, and the 
                       large SAURON $+$ ATLAS$^{\rm 3D}$ teams address many subjects that are beyond the scope of
                       studies by our team or by the Nuker team.  There is also one difference in analysis that
                       makes me uncomfortable -- in their work, they generally do not decompose galaxies into bulge
                       and disk parts.  {\it It is therefore all the more remarkable that careful work without using
                       component decomposition and our work that always is based on component decomposition
                       converge on pictures that are so similar.}  E.{\ts}g., the separate parameter correlations
                       for bulges and disks that are shown here in Figure~4 are visible as \hbox{pure-bulge~and~pure-disk} 
                       boundaries of parameter correlation {\it regions\/} shown in Cappellari \etal (2013b).
                       In their diagrams, the parameter space between our bulge and disk correlations is filled in
                       with intermediate-Hubble-type galaxies that have $0 < B/T < 1$.  Similarly, Cappellari \etal
                       (2011) and Kormendy \& Bender (2012) both revive the ``parallel sequence'' galaxy 
                       classification of van den Bergh (1976), as do Laurikainen \etal (2011).
                       Kormendy and Bender (2012) also add Sph galaxies (as distinct from ellipticals) to the classification.  
                      } \vskip 1pt

\item[\phantom{(0)}{\kern -3pt}]{What may appear as a difference between Section 4 and the SAURON $+$ ATLAS$^{\rm 3D}$ 
                       work is our emphasis on many E{\ts}--{\ts}E dichotomy classification criteria versus their 
                       distinction based only on fast versus slow rotation.  However, Lauer (2012) shows that the SAURON $+$
                       ATLAS$^{\rm 3D}$ division into fast and slow rotators is essentially equivalent to the 
                       division between coreless and core galaxies.  The equivalence is not exact based in the rotation 
                       amplitude parameter $\lambda_{r_e/2}$ (within 1/2 of the effective radius $r_e$) chosen by the
                       SAURON and ATLAS$^{\rm 3D}$ teams.  But it becomes much more nearly exact if slow and fast 
                       rotators are divided at a slighly higher rotation rate, $\lambda_{r_e/2} = 0.25$.  In unpublished
                       work, I found an essentially equivalent result for the original SAURON kinematic classification,
                       in which slow rotators have $\lambda_R < 0.1$ and fast rotators have $\lambda_R > 0.1$ as defined
                       in Emsellem \etal (2007).  If the division is instead made at $\lambda_R = 0.175$, then core and
                       coreless ellipticals are separated essentially perfectly.  (The only exception in KFCB is NGC 4458, 
                       which is slowly rotating but coreless.  But it is almost exactly round, and rotating galaxies that
                       are seen face-on will naturally look like slow rotators.)  The more nuanced ATLAS$^{\rm 3D}$ look 
                       at elliptical galaxy dynamics leads to a 
                       revised suggestion that fast and slow rotators should be separated at 
                       $\lambda_{r_e/2} = (0.265 \pm 0.01) \times \sqrt{\epsilon_{\rm e/2}}$
                       (Emsellem \etal 2011, Equation 4).  A typical $\epsilon = 0.2$ for core-boxy galaxies and
                       $\epsilon = 0.35$ for coreless-disky galaxies (from Tremblay \& Merritt 1996)
                       then implies a division at $\lambda_{r_e/2} = 0.16$ and 0.12, respectively.  The typical intrinsic
                       ellipticaly of 0.4 found by Sandage, Freeman, \& Stokes (1970) for all ellipticals implies
                       $\lambda_{r_e/2} = 0.17$.  These values are closer to the rotation parameters 0.25 and 0.175 that 
                       divide core and coreless galaxies as found by Lauer (2012) and by my work, respectively.} \vskip 1pt 

\item[\phantom{(0)}{\kern -3pt}]{I suggest that the best way to divide
                       slow rotators from fast rotators is not to pick some arbitrary value of the rotation parameter
                       but rather to ask the galaxies what value of the rotation parameter produces the cleanest 
                       distinction into two kinds of galaxies as summarized in Section 4.1.1.  When this is done, the
                       E{\ts}--{\ts}E dichotomy as discussed in this paper and the large body of work done by the SAURON 
                       and ATLAS$^{\rm 3D}$ teams are remarkably consistent.
                      }\vskip 1pt

\item[(2$+$3 redux){\kern -3pt}]{A partial exception to the above conclusion is some of the $n$-body simulation work, e.{\ts}g.,
                       by Naab \etal (2014).  They  acknowledge the importance of major mergers in some ways that
                       are consistent with the story advocated~in~this~paper.  But their conclusion that ``The galaxies
                       most consistent with the class of non-rotating round early-type galaxies grow by gas-poor {\it minor\/}
                       mergers alone'' (emphasis added) is at best uncomfortable within the picture presented here.  The
                       core-boxy-nonrotating galaxies have a large range of mostly homogeneous properties with respect
                       to which the round ones do not stand out as different (e.{\ts}g., KFCB).  In particular, our
                       understanding of cores -- especially the tight correlations between core properties and BH masses --
                       depends on our picture that cores are scoured by black hole binaries that are formed in major 
                       mergers (see KFCB and Kormendy \& Bender 2009 for both the data and a review).  At best, {\it it remains
                       to be demonstrated that minor mergers -- which necessarily involve many small galaxies with}
                       (from Figure 7) {\it undermassive BHs -- can produce the very large BH masses and cores
                       that are seen in giant core ellipticals.  Dry minor mergers cannot do better than to preserve
                       the $M_\bullet/M_{\rm host}$ mass ratio.  Also, if many minor mergers are necessary{\ts}--{\ts}and
                       these galaxies are so massive that very many minor mergers are necessary to grow them{\ts}--{\ts}then
                       there is a danger of producing a central cluster of low-mass BHs that is never observed
                       as a cluster of compact radio sources and that is inherently unstable to the ejection of 
                       objects in small-$n$ $n$-body systems} (see KH13, p.~634).
                      }

\item[(11){\kern -3pt}]{I conclude with two sociological points: It is worth emphasizing that galaxy evolution work 
                        did not start in the 2000s.  Many results that were derived in the 1960s -- 1990s remain valid today.  
                        We should not forget them.  We should integrate them into our current picture of galaxy evolution.}

\item[(12){\kern -3pt}]{And finally:  Galaxy evolution work has changed 
                       profoundly in the SDSS and HST eras.  Before the early
1990s, {\it our goal was to understand the evolution of galaxy structure.}  Now, most emphasis on galaxy
structure has disappeared.  {\it Now, our goal is to understand the history of star formation in the 
universe.}  The main reason for this change is the common ground found between SDSS studies
of many thousands of galaxies and HST studies of very distant galaxies.  Necessarily, both kinds
of studies concentrate on galaxies whose images are a few arcsec across.  We do not resolve
structural details.  Mainly, we measure colors and magnitudes.  So galaxy evolution has
evolved into the study of the red sequence and blue cloud in the color-magnitude relation.
Star formation and its quenching are, of course, important.  But it would be enormously healthy 
if we could improve the dialog between SDSS$+$HST people and those -- such as this author -- who work 
on nearby galaxies whose star formation histories and structures can be studied in great detail.  
Conselice (2014) is an example of a paper that tries to bridge the gap.  We would benefit greatly
if we could completely connect the two approaches to galaxy evolution.
}

\end{enumerate}

\begin{acknowledgement}
Many of my ideas about galaxy evolution were forged in intense and enjoyable collaborations 
with Ralf Bender, Luis Ho, and the Nuker team.  It is a pleasure to thank all these people and many
more who I do not have room to list for fruitful conversations over many years.  I am especially 
grateful to Reinhard Genzel for stimulating and insightful discussions and to
Ralf Bender,
Dimitri Gadotti, and 
Eija Laurikainen 
for very helpful comments on this paper.
Any errors of interpretation that remain are of course my responsibility.  I thank 
Steve Allen,
Xinyu Dai,
Ying-Jie Peng, and
Simon Lilly for permission to copy figures.  My work on this paper
was supported by the Curtis T.~Vaughan, Jr.~Centennial Chair in Astronomy at the University of Texas.
\end{acknowledgement}

\section{References}

\frenchspacing

\nhi Allen, S. W., Evrard, A. E., \& Mantz, A. B. 2011, ARA\&A. 49, 409

\nhi Armitage, P. J., \& Natarajan, P. 2002, ApJ, 567, L9

\nhi Armitage, P. J., \& Natarajan, P. 2005, ApJ, 634, 921

\nhi Athanassoula, E., 2005, MNRAS, 358, 1477

\nhi Athanassoula, E. 2015, in Galactic Bulges, ed. E. Laurikainen, R. F. Peletier, \& D. A. Gadotti (New York: Springer), 
     in press (arXiv:1503.04804)

\nhi Baes, M., Buyle, P., Hau, G. K. T., \& Dejonghe, H. 2003, MNRAS, 341, L44

\nhi Baggett, W. E., Baggett, S. M., \& Anderson, K. S. J. 1998, AJ, 116, 1626

\nhi Barth, A. J., Greene, J. E., \& Ho, L. C. 2008, AJ,  136, 1179

\nhi Barth, A. J., Ho, L. C., Rutledge, R. E., \& Sargent, W. L. W. 2004, ApJ, 607, 90

\nhi Begelman, M.~C., Blandford, R.~D., \& Rees, M.~J,~1980, Nature, 287, 307

\nhi Behroozi, P. S., Wechsler, R. H., \& Conroy, C. 2013, ApJ, 770, 57

\nhi Bender, R.~1988, A\&A, 193, L7 

\nhi Bender, R., Burstein, D., \& Faber, S.~M.~1992, ApJ, 399, 462

\nhi Bender, R., Surma, P., D\"obereiner, S., M\"ollenhoff, C., \& Madejsky, R. 1989, A\&A, 217, 35

\nhi Best, P. N. 2006, Paper Presented at the Workshop on The Role of Black Holes in Galaxy Formation and Evolution, Potsdam, Germany, 
     2006 September 10\ts--\ts13 (see Cattaneo \etal 2009)

\nhi Best, P. N. 2007a, in ASP Conf. Ser. 379, Cosmic Frontiers, ed. M. Metcalfe \& T. Shanks (San Francisco, CA: ASP), 213

\nhi Best, P. N. 2007b, New Astron. Rev., 51, 168

\nhi Best,{\ts}P.{\ts}N.,{\ts}Kaiser,{\ts}C.{\ts}R.,{\ts}Heckman,{\ts}T.{\ts}M.,\ts\&{\ts}Kauffmann,{\ts}G.\ts2006,{\ts}MNRAS,\ts368,{\ts}L67

\nhi Birnboim, Y., \& Dekel, A. 2003, MNRAS, 345, 349

\nhi Blanton, M. R., \& Moustakas, J. 2009, ARA\&A, 47, 159

\nhi Bluck, A. F. L., Conselice, C. J., Bouwens, R. J., \etal 2009, MNRAS, 394, 51

\nhi Bluck, A. F. L., Conselice, C. J., Buitrago, F., \etal 2012, ApJ, 747, 34

\nhi Booth, C. M., \& Schaye, J. 2010, MNRAS, 405, L1

\nhi Bothun, G. D., \& Thompson, I. B. 1988, AJ, 96, 877

\nhi Bournaud, F. 2015, in Galactic Bulges, ed. E. Laurikainen, R. F. Peletier, \& D. A. Gadotti (New York: Springer), in press (arXiv:1503.07660)

\nhi Bournaud, F., Elmegreen, B.~G., \& Elmegreen, D.~M.~2007, ApJ, 670, 237     

\nhi Boylan-Kolchin, M., Ma, C.-P., \& Quataert, E. 2006, MNRAS, 369, 1081

\nhi Brooks, A., \& Christensen, C. 2015, in Galactic Bulges, ed. E. Laurikainen, R. F. Peletier, \& D. A. Gadotti 
                                             (New York: Springer), in press

\nhi Cacciato, M., Dekel, A., \& Genel, S. 2012, MNRAS, 421, 818

\nhi Cappellari, M. 2015, ARA\&A, in preparation

\nhi Cappellari, M., Emsellem, E., Bacon, R., \etal 2007, MNRAS, 379, 418 

\nhi Cappellari, M., Emsellem, E., Krajnovi\'c, D., \etal 2011, MNRAS, 416, 1680

\nhi Cappellari, M., McDermid, R. M., Alatalo, K., \etal 2013b, MNRAS, 432, 1862

\nhi Cappellari, M., Scott, N., Alatalo, K., \etal 2013a, MNRAS, 432, 1709

\nhi Cattaneo, A., Dekel, A., Devriendt, J., Guiderdoni, B., \& Blaizot, J. 2006, MNRAS, 370, 1651

\nhi Cattaneo, A., Dekel, A., Faber, S. M., \& Guiderdoni, B. 2008, MNRAS, 389, 567

\nhi Cattaneo, A., Faber, S. M., Binney, J., \etal 2009, Nature, 460, 213

\nhi Chiboucas, K., Karachentsev, I. D., \& Tully, R. B. 2009, AJ, 137, 3009

\nhi Ceverino, D., Dekel, A., \& Bournaud, F. 2010, MNRAS, 404, 2151

\nhi Ceverino, D., Dekel, A., Tweed, D., \& Primack, J. 2015, MNRAS, 447, 3291

\nhi Cole, D. R., \& Debattista, V. P. 2015, in Galactic Bulges, ed. E. Laurikainen, R. F. Peletier, \& D. A. Gadotti 
                                                (New York: Springer), in press 

\nhi Combes, F. 2015, in Galactic Bulges, ed. E. Laurikainen, R. F. Peletier, \& D. A. Gadotti (New York: Springer), in press (arXiv:1501.03603)

\nhi Combes, F., \& Sanders, R.~H.~1981, A\&A, 96, 164

\nhi Conselice, C. J. 2014, ARA\&A, 52, 291

\nhi Conselice, C. J., Yang, C., \& Bluck, A. F. L. 2009, MNRAS, 394, 1956

\nhi Courteau, S., Dutton, A. A., van den Bosch, F. C., et al. 2007, ApJ, 671, 203

\nhi Cuadra, J., Armitage, P. J., Alexander, R. D., \& Begelman, M. C. 2009, MNRAS, 393, 1423

\nhi Daddi, E., Dickinson, M., Morrison, G., \etal 2007, ApJ, 670, 156

\nhi Dai, X, Bregman, J. N., Kochanek, C. S., \& Rasia, E. 2010, ApJ, 719, 119

\nhi Davies, R.~L., Efstathiou, G., Fall, S.~M., Illingworth, G., \& Schechter, P.~L. 1983, ApJ, 266, 41

\nhi Dekel, A., \& Birnboim, Y. 2006, MNRAS, 368, 39

\nhi Dekel, A., \& Birnboim, Y. 2008, MNRAS, 383, 119

\nhi Dekel, A., Birnboim, Y., Engel, G., \etal 2009, Nature, 457, 451

\nhi Dekel, A., Sari, R., \& Ceverino, D. 2009, ApJ, 703, 785

\nhi de Vaucouleurs, G., de Vaucouleurs, A., \& Corwin, J. R. 1976, Second Reference Catalogue of Bright Galaxies (Austin, TX:
     University of Texas Press)

\nhi Dong, X.-B., Ho, L. C., Yuan, W., \etal 2012, ApJ, 755, 167

\nhi Dotti, M., Colpi, M., Haardt, F., \& Mayer, L. 2007, MNRAS, 379, 956

\nhi Dressler, A. 1980, ApJ, 236, 351

\nhi Ebisuzaki, T., Makino, J., Okamura, S.~K.~1991, Nature, 354, 212

\nhi Elbaz, D., Daddi, E., Le Borgne, D., \etal 2007, A\&A, 468, 33

\nhi Ellis, S. C., \& O'Sullivan, E. 2006, MNRAS, 367, 627


\nhi Elmegreen, B.~G., Bournaud, F., \& Elmegreen, D.~M.~2008, ApJ, 688, 67  

\nhi Elmegreen, B.~G., \& Elmegreen, D.~M.~2005, ApJ, 627, 632

\nhi Elmegreen, B.~G., Elmegreen, D.~M, Fernandez, M.~X., \& Lemonias, J.~J. 2009a, ApJ, 692, 12



\nhi Elmegreen,{\ts}D.{\ts}M.,{\ts}Elmegreen,{\ts}B.{\ts}G,{\ts}Marcus,{\ts}M.{\ts}T.,\ts\etal2009b,{\ts}ApJ,{\ts}701,{\ts}306

\nhi Elmegreen, D.~M., Elmegreen, B.~G, Ravindranath, S., \& Coe, D.~A.~2007, ApJ, 658, 763

\nhi Emsellem, E., Cappellari, M., Krajnovi\'c, D., et al.~2007, MNRAS, 379, 401 

\nhi Emsellem, E., Cappellari, M., Krajnovi\'c, D., \etal 2011, MNRAS, 414, 888

\nhi Escala, A., \& Del Valle, L. 2011, Int. J. Mod. Phys. E, 20, 79

\nhi Escala, A., Larson, R. B., Coppi, P. S., \& Mardones, D. 2004, ApJ, 607, 765

\nhi Escala, A., Larson, R. B., Coppi, P. S., \& Mardones, D. 2005, ApJ, 630, 152

\nhi Faber, S. M., \& Jackson, R. E. 1976, ApJ, 204, 668

\nhi Faber, S.~M., Tremaine, S., Ajhar, E. A., \etal 1997, AJ, 114, 1771

\nhi Faber, S. M., Willmer, C. N. A., Wolf, C., \etal 2007, ApJ, 665, 265

\nhi Fabian, A. C. 1994, ARA\&A, 32, 277  

\nhi Fabian, A. C. 2012, ARA\&A, 50, 455  

\nhi Falc\'on-Barroso, J. 2015, in Galactic Bulges, ed. E. Laurikainen, R. F. Peletier, \& D. A. Gadotti 
                                   (New York: Springer), in press (arXiv:1503.04590)

\nhi Falc\'on-Barroso, J., \& Knapen, J. H. 2012, eds., Canary Islands Winter School of Astrophysics, Vol XXIII, 
                                    Secular Evolution of Galaxies (Cambridge: Cambridge University Press)

\nhi Fall, S. M., \& Romanowsky, A. J. 2013, ApJ, 769, L26

\nhi Ferrarese, L. 2002, ApJ, 578, 90

\nhi Ferrarese, L., C\^ot\'e, P., Jord\'an, A., et al. 2006, ApJS, 164, 334

\nhi Filippenko, A. V., \& Ho, L. C. 2003, ApJ, 588, L13   

\nhi Filippenko, A. V., Ho, L. C., \& Sargent W. L. W. 1993, ApJ, 410, L75

\nhi Finlator, K., \& Dav\'e, R. 2008, MNRAS, 385, 2181

\nhi Fisher, D.~B., \& Drory, N.~2008, AJ, 136, 773


\nhi Fisher, D.~B., \& Drory, N.~2011, ApJ, 733, L47					

\nhi Fisher, D.~B., \& Drory, N. 2015, in Galactic Bulges, ed. E. Laurikainen, R. F. Peletier, \& D. A. Gadotti 
                                   (New York: Springer), in press


\nhi Forbes, J. C., Krumholz, M. R., Burkert, A., \& Dekel, A. 2014, MNRAS, 438, 1552

\nhi F\"orster Schreiber, N.~M., Genzel, R., Bouch\'e, N., \etal 2009, ApJ, 706, 1364

\nhi F\"orster Schreiber, N.~M., Genzel, R., Newman, S. F., \etal 2014, ApJ, 787, 38

\nhi F\"orster Schreiber, N.~M., Shapley, A. E., Erb, D. K., \etal 2011a, ApJ, 731, 65

\nhi F\"orster Schreiber, N.~M., Shapley, A. E., Genzel, R., \etal 2011b, ApJ, 739, 45

\nhi Gabor, J. M., \& Dav\'e, R. 2015, MNRAS, 447, 374

\nhi Gadotti, D. A. 2008, Mem. Soc. Astron. Ital., 75, 1 (arXiv:1208.2295)

\nhi Gadotti, D. A. 2009, MNRAS, 393, 1531

\nhi Gadotti, D. A. 2012, in IAU Symposium 295, The Intriguing Life of Massive Galaxies,
     ed. D. Thomas, A. Pasquali, \& I. Ferreras (Cambridge: Cambridge Univ. Press), 232

\nhi Gaskell, C. M. 2010, in The First Stars and Galaxies: Challenges for the Next Decade, 
     ed. D. J. Whalen, V. Bromm, \& N. Yoshida (Melville, NY: AIP), 261

\nhi Gaskell, C. M. 2011, in SF2A-2011: Proceedings of the Annual Meeting of the French Society 
     of Astronomy and Astrophysics, 
     ed. G. Alecian, K. Belkacem, R. Samadi, \& D. Valls-Gabaud (Paris: SF2A), 577

\nhi Gavazzi, G., Donati, A., Cucciati, O., et al. 2005, A\&A, 430, 411

\nhi Gavazzi, G., Franzetti, P., Scodeggio, M., Boselli, A., \& Pierini, D. 2000, A\&A, 361, 863

\nhi Gebhardt, K., Lauer, T. R., Kormendy, J., et al. 2001, AJ, 122, 2469

\nhi Gebhardt, K., Richstone, D., Ajhar, E. A., et al.~1996, AJ, 112, 105 

\nhi Genel, S., Naab, T., Genzel, R., \etal 2012, ApJ, 745, 11                             

\nhi Genzel, R., Burkert, A., Bouch\'e, N., \etal 2008, ApJ, 687, 59

\nhi Genzel, R., F\"orster Schreiber, N. M., Rosario, D., \etal 2014, ApJ, 796, 7

\nhi Genzel, R., Newman, S., Jones, T., \etal 2011, ApJ, 733, 101

\nhi Genzel, R., Tacconi, L.~J., Eisenhauer, F., \etal 2006, Nature, 442, 786

\nhi Genzel, R., Tacconi, L.~J., Rigopoulou, D., Lutz, D., \& Tecza, M.~2001, ApJ, 563, 527

\nhi Gonzalez, O. A., \& Gadotti, D. A. 2015, in Galactic Bulges, ed. E. Laurikainen, R. F. Peletier, \& D. A. Gadotti 
                                   (New York: Springer), in press (arXiv:1503.07252)

\nhi Gould, A., \& Rix, H.-W. 2000, ApJ, 532, L29

\nhi Governato, F., Brook, C., Mayer, L., \etal 2010, Nature, 463, 203

\nhi Graham, A. W. 2011, Discussion comment following Kormendy, Bender, \& Cornell 2011

\nhi Graham, A. W. 2015, in Galactic Bulges, ed. E. Laurikainen, R. F. Peletier, \& D. A. Gadotti (New York: Springer), in press (arXiv:1501.02937)

\nhi Graham, A. W., \& Scott, N. 2013, ApJ, 764, 151

\nhi Graham, A. W., \& Scott, N. 2015, ApJ, 798, 54

\nhi Greene, J. E., \& Ho, L. C. 2004, ApJ, 610, 722

\nhi Greene, J. E., \& Ho, L. C. 2007, ApJ, 670, 92   

\nhi Greene, J. E., Peng, C. Y., Kim, M., et al. 2010, ApJ, 721, 26

\nhi Hayasaki, K. 2009, PASJ, 61, 65

\nhi Heckman, T. M., \& Best, P. N. 2014, ARA\&A, 52, 589

\nhi Hibbard, J.~E., Guhathakurta, P., van Gorkom, J.~H., \& Schweizer, F.~1994, AJ, 107, 67

\nhi Hibbard, J.~E., \& Mihos, J.~C.~1995, AJ, 110, 140

\nhi Hibbard, J.~E., van der Hulst, J.~M., Barnes, J.~E., \& Rich, R.~M.~2001a, AJ, 122, 2969

\nhi Hibbard, J.~E., \& van Gorkom, J.~H. 1996, AJ, 111, 655

\nhi Hibbard, J.~E., van Gorkom, J.~H., Rupen, M.~P., \& Schiminovich, D.~2001b, in ASP Conference Series, Vol.~240,
       Gas and Galaxy Evolution, ed. J. E. Hibbard, M. P. Rupen, \& J. H. van Gorkom (San Francisco: ASP), 659

\nhi Hilz, M., Naab, T., Ostriker, J. P., et al. 2012, MNRAS, 425, 3119

\nhi Hinshaw, G., Larson, D., Komatsu, E., \etal 2013, ApJS, 208, 19         

\nhi Hirschmann, M., Khochfar, S., Burkert, A., \etal 2010, MNRAS, 407, 1016 

\nhi Ho, L. C. 2008, ARA\&A, 46, 475

\nhi Ho, L. C., Greene, J. E., Filippenko, A. V., \& Sargent, W. L. W. 2009, ApJS, 183, 1  

\nhi Hopkins, P.~F., Cox, T.~J., Dutta, S.~N., Hernquist, L., Kormendy, J., \& Lauer, T.~R. 2009a, ApJS, 181, 135  

\nhi Hopkins, P.~F., Lauer, T.~R., Cox, T.~J., Hernquist, L., \& Kormendy, J.               2009b, ApJS, 181, 486  

\nhi Hu, J. 2008, MNRAS, 386, 2242                     

\nhi Irwin, M. J., Belokurov, V., Evans, N. W., et al. 2007, ApJ, 656, L13

\nhi Ivanov, P. B., Papaloizou, J. C. B., \& Polnarev, A. G. 1999, MNRAS, 307, 79

\nhi Jahnke, K., \& Macci\`o, A. V. 2011, ApJ, 734, 92  


\nhi Joseph, R. D., \& Wright, G. S. 1985, MNRAS, 214, 87

\nhi Karachentsev, I., Aparicio, A., \& Makarova, L. 1999, A\&A, 352, 363

\nhi Karim, A., Schinnerer, E., Mart\'\i nez-Sansigre, A., \etal 2011, ApJ, 730, 61

\nhi Kere\v s, D., Katz, N., Weinberg, D. H., \& Dav\'e, R. 2005, MNRAS, 363, 2

\nhi Khochfar, S., Emsellem, E., Serra, P., \etal 2011, MNRAS, 417, 845

\nhi Kirby, E. M., Jerjen, H., Ryder, S. D., \& Driver, S. P. 2008, AJ, 136, 1866

\nhi Knobel, C., Lilly, S. J., Woo, J., \& Kova\v c, K. 2015, ApJ, 800, 24

\nhi Komatsu, E., Dunkley, J., Nolta, M. R., et al. 2009, ApJS, 180, 330

\nhi Kormendy, J.~1979a, in Photometry, Kinematics and Dynamics of Galaxies, ed.~D.~S.~Evans 
                        (Austin: Dept.~of Astronomy, Univ.~of Texas at Austin), 341

\nhi Kormendy, J.~1979b, ApJ, 227, 714

\nhi Kormendy, J.~1981, in The Structure and Evolution of Normal Galaxies, ed. S. M. Fall \&
                        D. Lynden-Bell (Cambridge: Cambridge Univ.~Press), 85

\nhi Kormendy, J. 1982, in Twelfth Advanced Course of the Swiss Society of Astronomy and Astrophysics,
                          Morphology and Dynamics of Galaxies, ed. L. Martinet \& M. Mayor
                          (Sauverny: Geneva Observatory), 113

\nhi Kormendy, J. 1985, ApJ, 295, 73

\nhi Kormendy, J. 1987, in Nearly Normal Galaxies: From the Planck Time to the Present, ed. S. M. Faber (New York: Springer), 163

\nhi Kormendy, J.~1993, in IAU Symposium 153, Galactic Bulges, ed. H. Dejonghe \& H. J. Habing (Dordrecht: Kluwer), 209 

\nhi Kormendy, J. 1999, in Galaxy Dynamics: A Rutgers Symposium, ed. D. Merritt, J. A. Sellwood, \& M. Valluri (San Francisco, CA: ASP), 124

\nhi Kormendy, J.~2008, in IAU Symposium 245, Formation and Evolution of Galaxy Bulges, ed.~M.~Bureau,~E.~Athanassoula, \&
                        B.~Barbuy (Cambridge: Cambridge University Press), 107

\nhi Kormendy, J.~2009, in ASP Conference Series, Vol.~419, Galaxy Evolution: Emerging Insights and Future Challenges, ed. S. Jogee, I. Marinova,
              L. Hao, \& G. A. Blanc (San Francisco: ASP), 87 (arXiv:0812.0806)

\nhi Kormendy, J.~2012, in Canary Islands Winter School of Astrophysics, Vol.~XXIII,
         Secular Evolution of Galaxies, ed. J. Falc\'on-Barroso and J. H. Knapen (Cambridge: Cambridge University Press), 1

\nhi Kormendy, J.~2015, Highlights of Astronomy, 16, 316

\nhi Kormendy, J., \& Bender, R.~1996, ApJ, 464, L119

\nhi Kormendy, J., \& Bender, R. 2009, ApJ, 691, L142

\nhi Kormendy, J., \& Bender, R. 2011, Nature, 469, 377

\nhi Kormendy, J., \& Bender, R.~2012, ApJS, 198, 2 

\nhi Kormendy, J., \& Bender, R.~2013, ApJ, 769, L5

\nhi Kormendy, J., Bender, R., Ajhar, E. A., \etal 1996a, ApJ, 473, L91

\nhi Kormendy, J., Bender, R., \& Cornell, M. E. 2011, Nature, 469, 374

\nhi Kormendy, J., Bender, R., Richstone, D., \etal 1996b, ApJ, 459, L57

\nhi Kormendy, J., Dressler, A., Byun, Y.-I., et al.~1994, in ESO/OHP Workshop on Dwarf Galaxies, 
     ed.~G. Meylan \& P.~Prugniel (Garching: ESO), 147

\nhi Kormendy, J., Drory, N., Bender, R., \& Cornell, M.~E.~2010, ApJ, 723, 54

\nhi Kormendy, J., \& Fisher, D.~B.~2005, in The Ninth Texas-Mexico Conference on Astrophysics, ed. S. Torres-Peimbert \& G. MacAlpine,
     RevMexA\&A, Serie de Conferencias, 23, 101

\nhi Kormendy, J., \& Fisher, D. B.~2008, in ASP Conference Series, Vol. 396, Formation and Evolution of Galaxy Disks,
      ed.~J.~G.~Funes \& E.~M.~Corsini (San Francisco: ASP), 297

\nhi Kormendy, J., Fisher, D.~B., Cornell, M.~E., \& Bender, R.~2009, ApJS, 182, 216 (KFCB)

\nhi Kormendy, J., \& Freeman, K. C., 2015, ApJ, submitted (arXiv:1411.2170)

\nhi Kormendy, J., \& Gebhardt, K. 2001, in 20$^{\rm th}$ Texas Symposium on Relativistic Astrophysics, 
     ed. J. C. Wheeler \& H. Martel (Melville, NY: AIP), 363

\nhi Kormendy, J., \& Ho, L.~C.~2013, ARA\&A, 51, 511

\nhi Kormendy, J., \& Kennicutt, R.~C.~2004, ARA\&A, 42, 603


\nhi Kravtsov, A. V., \& Borgani, S. 2012, ARA\&A, 50, 353

\nhi Kuntschner, H., Emsellem, E., Bacon, R., \etal 2010, MNRAS, 408, 97

\nhi Lang, P., Wuyts, S., Somerville, R. S., \etal 2014, ApJ, 788, 11

\nhi Lauer, T. R. 2012, ApJ, 759, 64

\nhi Lauer, T.~R., Ajhar, E. A., Byun, Y.-I., \etal 1995, AJ, 110, 2622 

\nhi Lauer, T.~R., Faber, S. M., Gebhardt, K., \etal 2005, AJ, 129, 2138

\nhi Lauer, T.~R., Faber, S. M., Richstone, D., \etal 2007a, ApJ, 662, 808  

\nhi Lauer, T.~R., Gebhardt, K., Faber, S. M., \etal 2007b, ApJ, 664, 226   

\nhi Laurikainen, E., \& Salo, H. 2015, in Galactic Bulges, ed. E. Laurikainen, R. F. Peletier, \& D. A. Gadotti 
                                   (New York: Springer), in press 

\nhi Laurikainen, E., Salo, H., Buta, R., \& Knapen, J. H. 2011, MNRAS, 418, 1452

\nhi Laurikainen, E., Salo, H., Buta, R., Knapen, J. H., \& Comer\'on, S. 2010, MNRAS, 405, 1089

\nhi Le Floc'h, E., Papovich, C., Dole, H. \etal 2005, ApJ, 632, 169

\nhi Lilly, S. J., Le F\`evre, O., Renzini, A., \etal 2007, ApJS, 172, 70

\nhi L\'opez-Sanjuan, C., Le F\`evre, O., Ilbert, O., \etal 2013, in ASP Conference Series, Vol. 477,
     Galaxy Mergers in an Evolving Universe, ed. W.-H. Sun, K. Xu, N. Scoville, \& D. Sanders (San Francisco: ASP), 159

\nhi Madore, B. F. 2015, in Galactic Bulges, ed. E. Laurikainen, R. F. Peletier, \& D. A. Gadotti (New York: Springer), in press 

\nhi Makarova, L. 1999, A\&AS, 139, 491

\nhi Makino, J., \& Ebisuzaki, T.~1996, ApJ, 465, 527

\nhi Marconi, A., \& Hunt, L. K.  2003, ApJ, 589, L21

\nhi Mateo, M. 1998, ARA\&A, 36, 435

\nhi Mayer, L. 2013, Class. Quantum Grav. 30, 244008

\nhi McConnell, N. J., \& Ma, C.-P. 2013, ApJ, 764, 184

\nhi McLure, R. J., \& Dunlop, J. S. 2002, MNRAS, 331, 795

\nhi McNamara, B. R., \& Nulsen, P. E. J. 2007, ARA\&A, 45, 117

\nhi M\'endez-Abreu, J. 2015, in Galactic Bulges, ed. E. Laurikainen, R. F. Peletier, \& D. A. Gadotti (New York: Springer), in press (1502.00265)

\nhi Merritt, D.~2006, ApJ, 648, 976

\nhi Merritt, D., \& Ferrarese, L. 2001, MNRAS, 320, L30  

\nhi Mihos, J. C., \& Hernquist, L. 1994, ApJ, 437, L47

\nhi Milosavljevi\'c, M., \& Merritt, D.~2001, ApJ, 563, 34

\nhi Milosavljevi\'c, M., Merritt, D., Rest, A., \& van den Bosch, F.~C.~2002, MNRAS, 331, L51

\nhi Naab, T., 2013, in IAU Symposium 295, The Intriguing Life of Massive Galaxies, ed. D. Thomas, A. Pasquali, \&
                     I. Ferreras (Cambridge:~Cambridge Univ.~Press), 340

\nhi Naab, T., Oser, L., Emsellem, E., \etal 2014, MNRAS, 444, 3357

\nhi Nieto, J.-L., Bender, R., \& Surma, P.~1991, A\&A, 244, L37 

\nhi Noeske, K. G., Weiner, B. J., Faber, S. M., \etal 2007, ApJ, 660, L43

\nhi Ostriker, J. P. 2006, Paper Presented at the Workshop on The Role of Black Holes in Galaxy Formation and Evolution,              
     Potsdam, Germany, 2006 September 10\ts--\ts13 (see Cattaneo \etal 2009)

\nhi O'Sullivan, E., Forbes, D. A., \& Ponman, T. J. 2001, MNRAS, 328, 461

\nhi Pellegrini, S. 1999, A\&A, 351, 487

\nhi Pellegrini, S. 2005, MNRAS, 364, 169

\nhi Peng, C. Y. 2007, ApJ, 671, 1098  

\nhi Peng, Y.-J., Lilly, S. J., Kova\v c, K., \etal 2010, ApJ, 721, 193

\nhi Peng, Y.-J., Lilly, S. J., Renzini, A., \& Carollo, M. 2012, ApJ, 757, 4

\nhi Peterson, B. M., Bentz, M. C., Desroches, L.-B., \etal 2005, ApJ, 632, 799; Erratum. 2006, ApJ, 641, 638 

\nhi Pildis, R. A., Schombert, J. M., \& Eder, J. A. 1997, ApJ, 481, 157

\nhi Planck Collaboration 2014, A\&A, 571, A16

\nhi Puech, M., Hammer, F., Rodrigues, M., \etal 2014, MNRAS, 443, L49

\nhi Quinlan, G. D., \& Hernquist, L. 1997, New Astron., 2, 533

\nhi Ravindranath, S., Ho, L.~C., Peng, C.~Y., Filippenko, A.~V., \& Sargent, W.~L.~W. 2001, AJ, 122, 653

\nhi Rees, M. J., \& Ostriker, J. P. 1977, MNRAS, 179, 541

\nhi Reines, A. E., Sivakoff, G. R., Johnson, K. E., \& Brogan, C. L. 2011, Nature, 470, 66  

\nhi Renzini, A.~1999, in The Formation of Galactic Bulges, ed.~C.~M.~Carollo, H.{\ts}C.{\ts}Ferguson \& R.{\ts}F.{\ts}G.{\ts}Wyse (Cambridge:
             Cambridge Univ. Press),\ts9

\nhi Rest, A., van den Bosch, F.~C., Jaffe, W., \etal ~2001, AJ, 121, 2431

\nhi Rodighiero, G., Daddi, E., Baronchelli, I., \etal 2011, ApJ, 739, L40

\nhi Romanowsky, A. J., \& Fall, S. M. 2012, ApJS, 203, 17

\nhi Rusli, S. P., Thomas, J., Saglia, R. P., \etal 2013, AJ, 146, 45

\nhi Salmi, F., Daddi, E., Elbaz, D., \etal 2012, ApJ, 754, L14

\nhi Salo, H., Laurikainen, E., Laine, J., \etal 2015, ApJ, accepted for publication

\nhi S\'anchez-Bl\'azquez, P. 2015, in Galactic Bulges, ed. E. Laurikainen, R. F. Peletier, \& D. A. Gadotti (New York: Springer), in press (1503.08105)

\nhi Sandage, A., Freeman, K.~C., \& Stokes, N.~R.~1970, ApJ, 160, 831

\nhi Sanders, D. B., Soifer, B. T., Elias, J. H., \etal 1988a, ApJ, 325, 74

\nhi Sanders, D. B., Soifer, B. T., Elias, J. H., Neugebauer, G., \& Matthews, K. 1988b, ApJ, 328, L35

\nhi Sani, E., Marconi, A., Hunt, L. K., \& Risaliti, G. 2011, MNRAS, 413, 1479

\nhi Schechter, P. 1976, ApJ, 203, 297

\nhi Schiminovich, D., Wyder, T. K., Martin, D. C., \etal 2007, ApJS, 173, 315

\nhi Schulze, A., \& Gebhardt, K. 2011, ApJ, 729, 21

\nhi  Schweizer, F. 1987, in Nearly Normal Galaxies: From the Planck Time to the Present, ed. S. M. Faber (New York: Springer-Verlag), 18

\nhi  Schweizer, F. 1990, in Dynamics and Interactions of Galaxies, ed. R. Wielen (New York: Springer-Verlag), 60

\nhi  Schweizer, F. 1998, in 26$^{\rm th}$ Advanced Course of the Swiss Society of Astronomy and Astrophysics, Galaxies: 
        Interactions and Induced Star Formation, ed. D. Friedli, L. Martinet, \& D. Pfenniger (New York: Springer-Verlag), 105

\nhi Sellwood, J. A. 2014, Rev. Mod. Phys., 86, 1

\nhi S\'ersic,{\ts}J.{\ts}L.\ts1968, Atlas{\ts}de{\ts}Galaxias{\ts}Australes (C\'ordoba:{\ts}Observatorio{\ts}Astron\'omico, Universidad de C\'ordoba)

\nhi Shen, J., \& Li, Z.-Y. 2015, in Galactic Bulges, ed. E. Laurikainen, R. F. Peletier, \& D. A. Gadotti (New York: Springer), in press

\nhi Shih, D. C., Iwasawa, K., \& Fabian, A. C. 2003, MNRAS, 341, 973

\nhi Speagle,{\ts}J.{\ts}S., Steinhardt,{\ts}C.{\ts}L., Capak,{\ts}P.{\ts}L., \& Silverman,{\ts}J.{\ts}D.\ts2014, ApJS, 214, 15


\nhi Tacconi, L.~J., Genzel, R., Neri, R., \etal 2010, Nature, 463, 781

\nhi Tacconi, L. J., Neri, R., Genzel, R., \etal 2013, ApJ, 768, 74

\nhi Tasca, L. A. M., Le F\`evre, O., L\'opez-Sanjuan, C., \etal 2014, A\&A, 565, A10

\nhi Thomas, D., Maraston, C., \& Bender, R. 2002a, Rev. Mod. Astron., 15, 219 (arXiv:astro-ph/0202166)

\nhi Thomas, D., Maraston, C., \& Bender, R. 2002b, Ap. Space Sci., 281, 371

\nhi Thomas, D., Maraston, C., Bender, R., \& Mendes de Oliveira, C.~2005, ApJ, 621, 673

\nhi Thornton, C. E., Barth, A. J., Ho, L. C., Rutledge, R. E., \& Greene, J. E. 2008, ApJ, 686, 892

\nhi Toomre, A.~1964, ApJ, 139, 1217  

\nhi Toomre A.~1977, in The Evolution of Galaxies and Stellar Populations, ed. B. M. Tinsley \& R. B. Larson (New Haven:
             Yale Univ.~Obs.), 401

\nhi Toomre, A., \& Toomre, J.~1972, ApJ, 178, 623

\nhi Tremaine, S., Gebhardt, K., Bender, R., \etal 2002, ApJ, 574, 740


\nhi Tremblay, B., \& Merritt, D.~1996, AJ, 111, 2243

\nhi van Albada, T.~S.~1982, MNRAS, 201, 939

\nhi van den Bergh, S. 1976, ApJ, 206, 883

\nhi Whitaker,{\ts}K.{\ts}E., van{\ts}Dokkum,{\ts}P.{\ts}G., Brammer,{\ts}G., \& Franx,{\ts}M.~2012, ApJ,~754,~L29

\nhi White, S.~D.~M., \& Rees, M.~J.~1978, MNRAS, 183, 341


\nhi Williams, M. J., Bureau, M., \& Cappellari, M. 2009, MNRAS, 400, 1665

\nhi Woo, J., Dekel, A., Faber, S. M., \& Koo, D. C. 2015, MNRAS, 448, 237

\nhi Wrobel, J. M., \& Ho, L. C. 2006, ApJ, 646, L95

\nhi Wuyts, S., F\"orster Schreiber, N. M., Genzel, R., \etal 2012, ApJ, 753, 114

\nhi Wuyts, S., F\"orster Schreiber, N. M., van der Wel, A., \etal 2011, ApJ, 742, 96

\nhi Zaritsky, D. 2015, in Galactic Bulges, ed. E. Laurikainen, R. F. Peletier, \& D. A. Gadotti (New York: Springer), in press

\vfill\eject

\end{document}